**Title:** BGnet: Accurate and rapid background estimation in single-molecule localization microscopy with deep neural nets

**Authors:** Leonhard Möckl[a,1], Anish R. Roy[a,1], Petar N. Petrov[a], and W.E. Moerner[a,2]
[a]Department of Chemistry, Stanford University, Stanford, CA 94305
[1]These authors contributed equally.
[2]Corresponding author: wmoerner@stanford.edu

**Abstract:** Background fluorescence, especially when it exhibits undesired spatial features, is a primary factor for reduced image quality in optical microscopy. Structured background is particularly detrimental when analyzing single-molecule images for 3D localization microscopy or single-molecule tracking. Here, we introduce BGnet, a deep neural network with a U-net-type architecture, as a general method to rapidly estimate the background underlying the image of a point source with excellent accuracy, even when point spread function (PSF) engineering is in use to create complex PSF shapes. We trained BGnet to extract the background from images of various PSFs and show that the identification is accurate for a wide range of different interfering background structures constructed from many spatial frequencies. Furthermore, we demonstrate that the obtained background-corrected PSF images, both for simulated and experimental data, lead to a substantial improvement in localization precision. Finally, we verify that structured background estimation with BGnet results in higher quality of super-resolution reconstructions of biological structures.

**Significance:** A main factor that degrades image quality in fluorescence microscopy is unwanted background fluorescence. Background is almost never uniform, especially in complex samples. Rather, background usually exhibits some structure, making it very difficult to distinguish from the signal of interest. Due to this challenging problem, background fluorescence is often assumed to be uniform even though it is not. This assumption leads to deteriorated image quality, e.g. in localization-based super-resolution microscopy, and to the introduction of uncharacterized biases. To overcome this challenge, we developed a general framework rooted in deep learning to accurately and rapidly estimate arbitrarily structured background using several distinct image shapes for the single emitter. Proper background estimation allows for critical performance improvement of various optical microscopy methods.

**Main Text:**
**Introduction**
In optical microscopy, the term "background" (BG) summarizes contributions to an image that do not arise from the species that is investigated, but from other sources.(1, 2) These contributions lower the quality of the image and are, therefore, unwanted. For example, when performing fluorescence microscopy of a cellular protein labeled via immunochemistry, antibodies may bind nonspecifically to other cellular components or to the sample chamber, or the sample itself can exhibit autofluorescence.(3)

Often, during camera-based localization within a small region of interest, the BG structure of an image is considered to be uniform within that region and is accounted for by subtraction of a mean or median fluorescence signal which is extracted from an image area that has no contribution from the fluorescently labeled species of interest.(4) The assumption of unstructured (uniform) BG is, however, an oversimplification in most situations. For example, in biological microscopy, a typical specimen such as a cell or a tissue slice features a huge number of different components that are distributed over many different spatial length scales which may be autofluorescent.(5) A fluorescent probe introduced to label a component may also bind nonspecifically to other components. Therefore, the resulting fluorescent BG will be composed of many different spatial frequencies. Thus, this type of BG can be termed "structured BG" (sBG).(6)

sBG is especially detrimental when single emitters such as single molecules are detected and imaged to estimate their position on the nanometer scale as is done in localization-based super-resolution microscopy methods (e.g. PALM, STORM, f-PALM) or single-molecule tracking.(7-9) In these approaches, a BG-free model function of the point spread function (PSF), i.e. the response function of the microscope when a single emitter is imaged, is fit to the experimentally recorded camera image of the single molecule containing BG.(2, 10) In the simplest case, the standard (open aperture) PSF of a typical microscope can be approximated by a 2D Gaussian. For 3D imaging, more complex PSFs have been developed via PSF engineering in the Fourier plane, and the information about z position is encoded in the more complex image.(11) Similar PSF engineering strategies can be used to encode other variables such as emitter orientation, wavelength, etc. (12-14)

While unstructured BG can be easily accounted for in the PSF fitting process as an additive offset, removing sBG is much more challenging: a simple subtraction of some number will just shift the average BG magnitude, but not remove the underlying structure. The remaining sBG changes the PSF shape which can strongly affect the result of the position estimation, regardless of the fitting algorithm used (e.g. least squares or maximum likelihood estimation, MLE).(15, 16)

Unfortunately, correction for sBG is not trivial as it can exhibit contributions from various spatial frequencies. Any approach to remove sBG must be able to differentiate between the spatial information from the PSF alone, which must be retained, and the spatial information of the sBG.(17, 18) A recent Bayesian approach estimated background for a specific case,(19) but more general background estimation procedures are needed. Methods such as sigma clipping (20, 21) have been developed to account for sBG; however, for more complex PSFs used in 3D imaging, sBG estimation with these approaches is very challenging. Therefore, even though sBG is a prominent feature for experimental datasets, the simple assumption of constant BG is still widely used today. In this work, we address this problem by employing advanced image analysis with deep neural networks (DNNs), using the network to extract the sBG for proper removal.

**Results and Discussion**
*General workflow and BGnet architecture*
Here, we introduce BGnet, a DNN that allows for rapid and accurate estimation of sBG. DNNs are versatile tools for various applications, among which image analysis for general purpose feature recognition as well as for optical microscopy are prominent.(22-26) Recently, the U-net architecture has been demonstrated to be well suited for image segmentation.(27, 28) Fundamentally, image segmentation is similar to sBG estimation: A feature – the PSF without BG – is overlaid with the sBG, which should be identified from the combined image in order to subsequently remove it. Therefore, we suspected that a U-net-type architecture might also be applicable for sBG estimation in optical microscopy, as schematically depicted in Fig. 1(a). The architecture of BGnet is depicted in Fig. 1(b), illustrating the U-shaped architecture of the network. The fundamental idea is to first condense the spatial size of the input image stepwise while increasing its filter space. Then, stepwise upsampling is performed until the original spatial scale of the image is obtained, and the filter space is reduced in turn. This is often termed encoder-decoder architecture.(22, 29) In U-net-type architectures, the output before each downsampling (left arm of the U) is concatenated with the result of the upsampling (right arm of the U) at corresponding spatial scales. This is reminiscent of residual nets where the output of a layer is added to the output of a deeper layer via skipped connections.(30)

First, we provided BGnet with training data that covers the wide parameter space that sBG estimation poses: A given PSF that should be analyzed can have various shapes and sizes at different axial positions of the emitter; and many different spatial frequencies can combine to form the sBG. Therefore, we turned to accurate PSF simulations of three commonly used PSFs: The standard open aperture (OA) PSF, the double-helix (DH) PSF with 2 µm axial range (15, 31), and the Tetrapod PSF with 6 µm axial range (Tetra6 PSF).(32) Also, we included an arbitrary PSF with a rather chaotic shape to test whether our approach is robust against PSF shapes that do not exhibit a well-defined structure. As a model for sBG, we chose *Perlin noise* because it is (i) able to accurately resemble sBG encountered under most experimental conditions and (ii) precisely controllable in its spatial frequency composition (see Fig. S1 for an overview).(33)

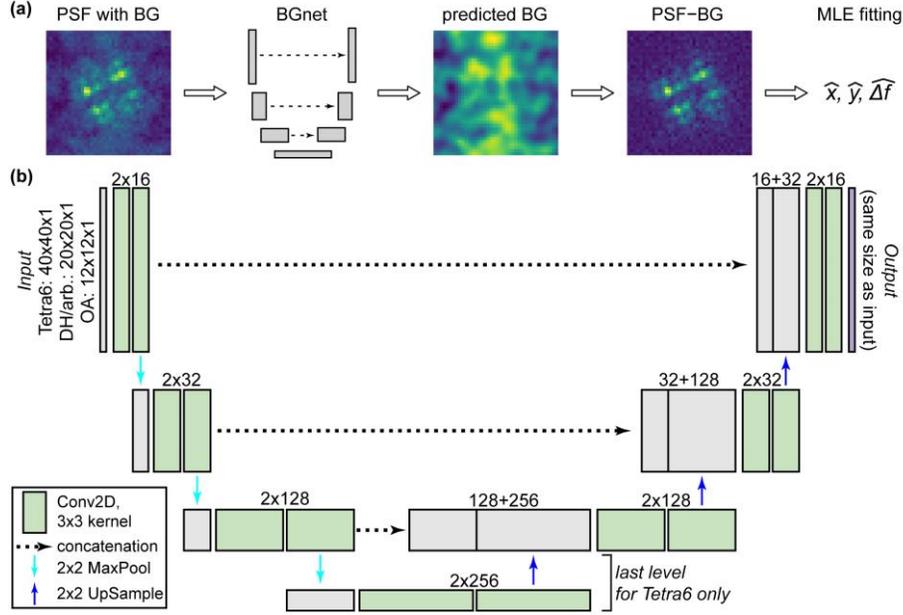

**Fig. 1:** General approach and BGnet architecture. (a) BGnet receives an image of a PSF (here the Tetra6 PSF) with BG. Its output is the predicted BG contribution at each pixel. Thus, the predicted BG can be readily subtracted from the input PSF image. The BG-corrected PSF can subsequently be analyzed, for example via MLE fitting for position estimation in x and y with defocus ∆f. (b) The PSF images are supplied to BGnet as single-channel 12x12 (OA PSF), 20x20 (DH, arbitrary PSF), or 40x40 pixel images. After two 2D convolutions with 16 filters, Batch Normalization and ReLu activation, 2x2 MaxPooling is performed. Two 2D convolutions with 32 filters are performed. The output is again subjected to 2x2 MaxPooling, followed by two 2D convolutions with 128 filters. An additional 2x2 MaxPooling, followed by two more 2D convolutional layers with 256 filters, is performed for the Tetra6 PSF only. The output of the 2D convolutional layers with the lowest spatial size is upsampled (2x2) and concatenated with the output of the 2D convolutional layer that was supplied to the final 2x2 MaxPooling. Upsampling, concatenation, and 2D convolution are repeated until the spatial scale of the image is again 12x12, 20x20, or 40x40, respectively. The last layer is a 12x12x1, 20x20x1, or 40x40x1 2D convolutional layer, returning the predicted BG.

PSFs were simulated by means of vectorial diffraction theory (34, 35) using simulation parameters matching typical experimental values and accurately characterized aberrations, determined via phase retrieval as previously published.(16) The PSFs were simulated at different focal positions and different distances away from a glass coverslip (n = 1.518) in water (n = 1.33) (see Table S1-S4 for simulation parameters). The Perlin noise used for sBG modelling contained spatial frequencies of $L/12$, $L/6$, $L/4$, and $L/2$ for the OA PSF; $L/20$, $L/10$, $L/5$, and $L/2$ for the DH and arbitrary PSF; and $L/40$, $L/20$, $L/10$, $L/5$, and $L/2$, for the Tetra6 PSF; with $L$ being the size of the image in pixels (12, 20, or 40, respectively). Notably, the contribution of each individual frequency was not restricted and ranges anywhere between 0 and 100%. Signal and BG photons were simulated across a wide range, dependent on the PSF, to generate training and validation data. Each input PSF was normalized between 0 and 1 and the target, i.e. the true BG that BGnet is trained to return, was scaled identically. Therefore, the BGnet not only predicts the structure of the BG, but also its intensity relative to the input PSF image at each pixel.

BGnet was implemented in Keras with Tensorflow backend and trained on a desktop PC equipped with 64 GB RAM, an Intel Xeon E5-1650 processor, and an Nvidia GeForce GTX Titan GPU. Convergence was reached after training for approx. one hour (OA PSF) to approx. nine hours (Tetra6 PSF). Detailed training parameters are listed in Table S5. All validation experiments were done with an independent dataset that was not part of the training dataset.

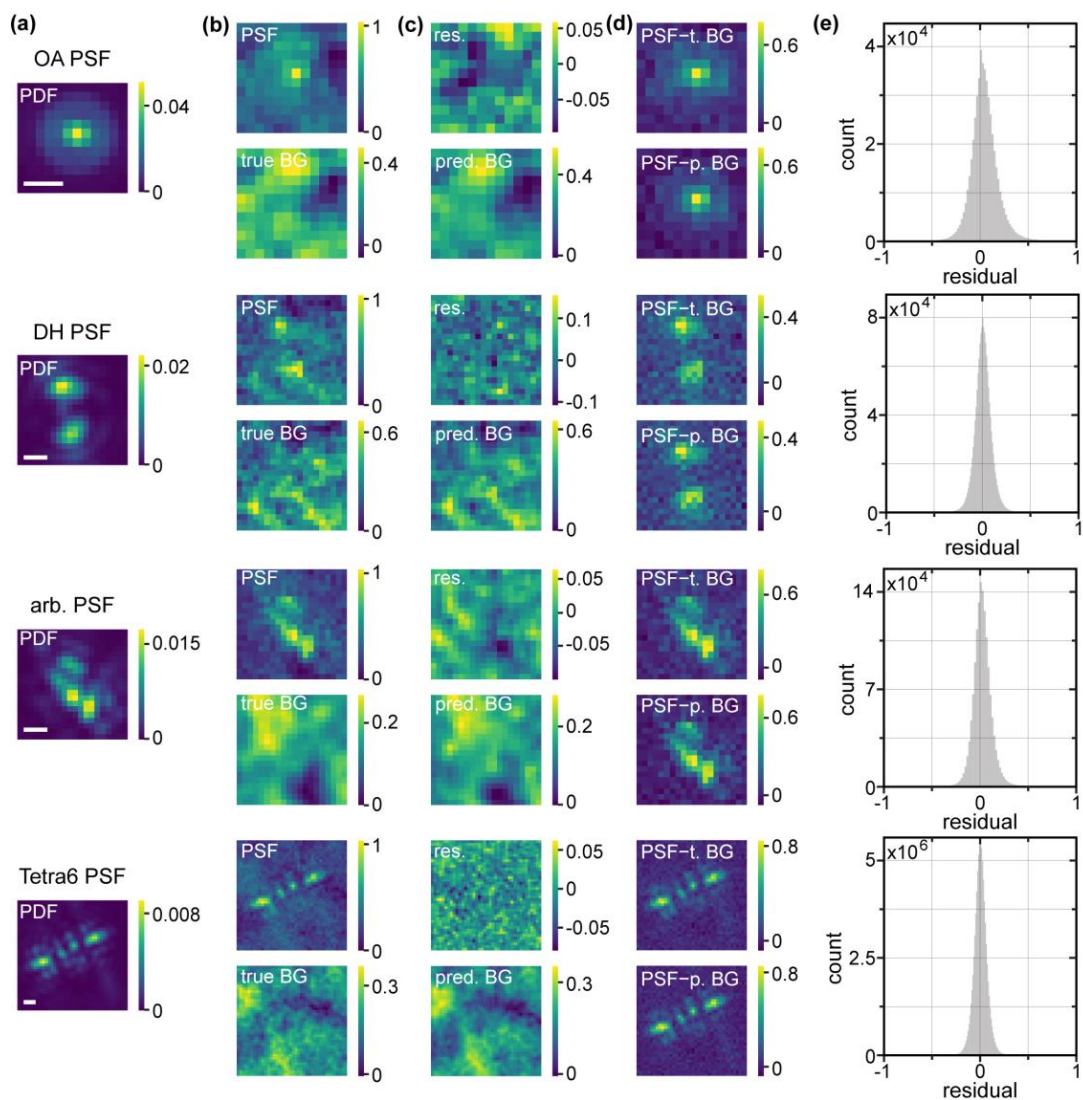

**Fig. 2:** Representative examples for BG estimation with BGnet and overall performance. (a) Example probability density functions (PDFs) for the four investigated PSFs. (b) The BG-corrupted input PSFs, normalized between 0 and 1, and the underlying (true) BGs. The signal

photon count for the depicted PSFs is 4723, 5275, 5994, and 37637 and the average BG photon count per pixel is 147, 137, 26, and 127 (from top to bottom). (c) The BG prediction by BGnet on the same intensity scale as the input PSF and the residual between true and predicted BG. (d) The original PSFs corrected for BG either using the true BG or the predicted BG. Note that negative pixel values for the BG-corrected PSFs are only a side effect of PSF normalization and originate from Poisson noise fluctuations. (e) Pixelwise residuals for all the PSFs of the validation dataset. For this analysis, true and predicted BG were scaled between 0 and 1, such that all residuals ranged from -1 to 1. The scale bar in (a) is 500 nm.

*BGnet accurately estimates sBG from images of various PSF shapes*
Fig. 2 shows representative examples for the PSF simulation process and the performance of BGnet on validation data. In Fig. 2(a), the probability density functions (PDFs) are shown as a reference for one axial position. The PSFs containing BG (Fig. 2(b) top images for each PSF) are supplied to BGnet, which returns the predicted BGs (Fig. 2(c) bottom images). The agreement between true (Fig. 2(b) bottom images) and predicted BGs is excellent, reflected in small residuals (Fig. 2(c) top images).The obtained BGs can then be subtracted from the PSF images for BG correction. The strongly improved quality of the PSF shapes after BG correction is evident. Illustrating the quality of the BG estimation, the images for the PSFs corrected with the true BGs and the PSF corrected with the predicted BGs are very similar (Fig. 2(d)). For additional representative examples, see Fig. S2-S5.

To quantify the overall agreement between true and predicted BGs, we normalized each pair of true and predicted BGs between 0 and 1 (otherwise, due to varying signal and background levels, the residuals cannot be directly compared). Then, we calculated the pixelwise difference between true and predicted BGs for all the PSFs in the validation dataset. The result is depicted in Figure 2(e). Clearly, the residuals, which can range between -1 and 1, form a narrow distribution which is centered at 0. This indicates that the BG is accurately estimated by BGnet. Importantly, this process is very fast. 3500-5000 PSFs were analyzed in 4 to 30 seconds on a standard desktop PC (quickest for the OA PSF, slowest for the Tetra6 PSF due to the different image sizes), which corresponds to approximately 1 to 6 ms/PSF, suitable for real-time analysis. Using a PC equipped with a dedicated GPU could speed up BG estimation even more if required.

*BGnet strongly improves localization precision of single molecules*
The good agreement between predicted and true BGs is promising. However, it is critical to verify that removing the predicted sBGs translates to improved precision of extracted single-molecule parameters compared to conventional BG correction approaches. Therefore, we explored how BG correction with BGnet affects the 3D emitter localization precision via MLE fitting of the images to the models (see Materials and Methods). For this analysis, we simulated PSFs at various distances from the coverslip and various focal positions. Furthermore, we varied the signal photons and the average BG photons per pixel over a wide range, specific to each PSF and used values typical for experiments, which resulted in 90 different parameter combinations for the OA, the DH, and the arbitrary PSF and in 270 different parameter combination for the Tetra6 PSF (see Table S6). Each parameter combination was realized 100 times with the respective PSF position held constant. However, each of the 100 PSF realizations for a specific parameter combination was corrupted with different BG structures. As the true PSF position is always the same, the "spread" of the localizations (i.e. the mean of the standard deviations of the position estimates in each spatial dimension x, y, and Δ$f$) directly reports on the effect of BG subtraction.

We analyzed four different scenarios: (i) BG correction with the predicted BG from BGnet; (ii) BG correction with the ground-truth, true BG, (iii) a BG-free PSF that only exhibits Poisson noise; and (iv) conventional BG correction with a constant BG as typically assumed. Case (iii) is a baseline reference which exhibits the best localization precision obtainable in a BG free scenario for the detected photons assumed. The results are depicted in Fig. 3.

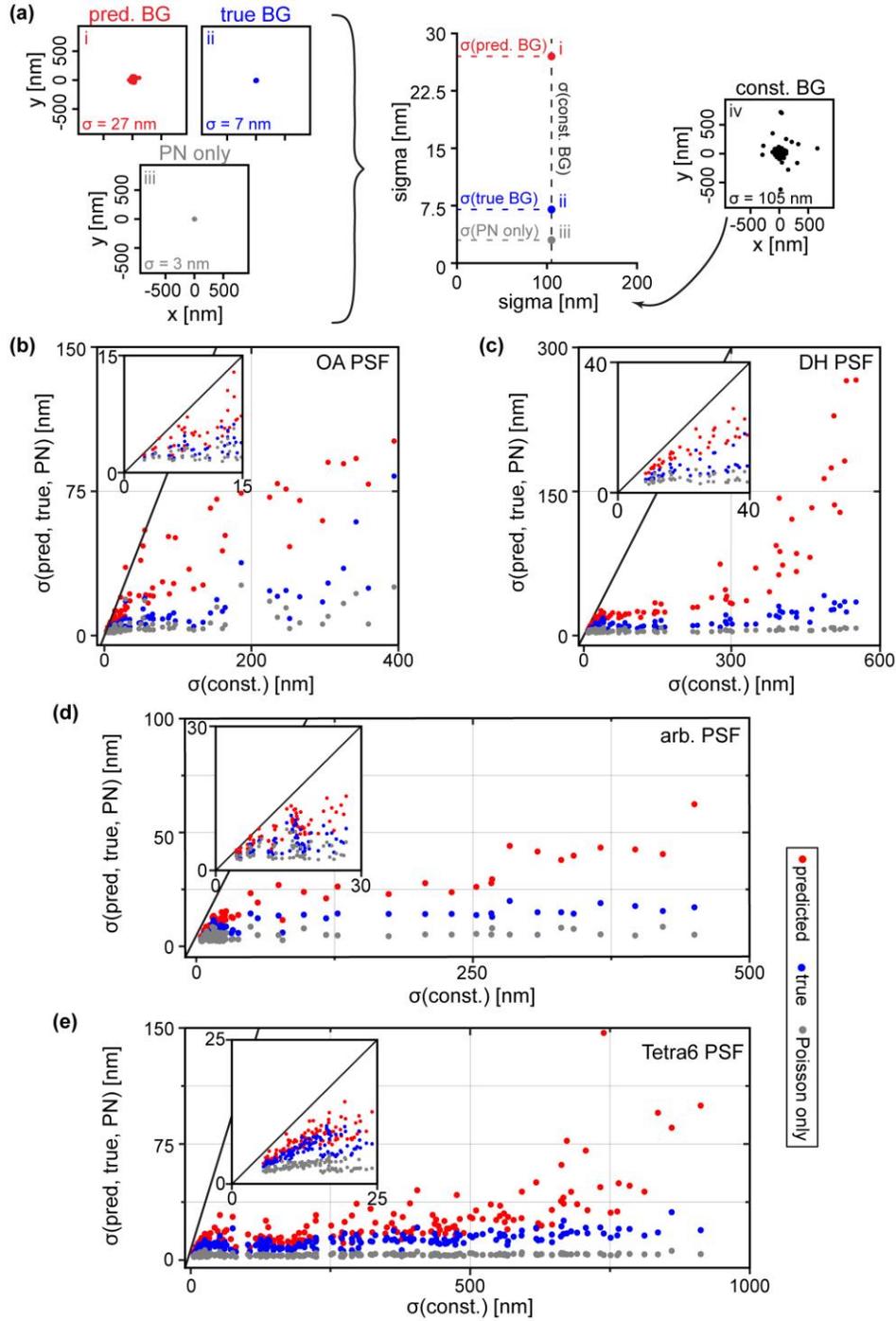

**Fig. 3:** Significant improvement in localization precision occurs with MLE fitting when BG correction with BGnet is used. (a) Schematic of data visualization approach and representative x/y scatter plots for a given parameter combination for the OA PSF. The spreads of the position estimates are 105, 27, 7, and 3 nm for scenarios (i) to (iv), respectively leading to the points placed on the plot in the center. (b) to (e) The spread of the position estimates for scenarios (i), (i), (iii) is plotted against the spread of the position estimates for scenario (iv), that is, the constant BG estimate is used as a reference. The inset shows magnifications. The gray line has a slope of unity and thus indicates equal performance. Points below that line perform better than the reference case (constant BG estimate). (b) OA PSF, (c) DH PSF, (d) arbitrary PSF, (e) Tetra6 PSF. For the OA PSF, only the x/y position estimates are considered; the other cases use 3D spreads.

For each of the four scenarios, the MLE fitting of the images with different background structures yields 100 position estimates, the spreads of which can be quantified by a standard deviation. The x/y scatter plots in Fig. 3(a) show a representative result for the OA PSF (10,000 signal photons, 150 average BG photons/pixel, emitter at 2 µm, focal position for scatter plot at 0.5 µm; for further examples of all investigated PSFs, including x/$\Delta f$ scatter plots, see Fig. S6 to S9). The spreads of the position estimates for scenarios (i), (ii), and (iii) are plotted against the spread of the reference scenario, the constant BG estimate (scenario (iv), on the right) for each parameter combination. Figure 3(b) to (e) depicts the OA, DH, arbitrary, and Tetra6 PSF, respectively. The significant improvement in localization precision when using BGnet is evident for all PSFs and any condition: The spread of the position estimates is much smaller when BG correction with BGnet is used. Nearly all points corresponding to BG correction with BGnet are located far below the line with slope unity. This demonstrates that the excellent accuracy with which BGnet extracts the BG from PSF images directly results in improved localization precision.

For many cases, the crude BG correction with a constant BG leads to spreads of hundreds of nanometers, which is considerably reduced when BG correction with BGnet is performed. These extreme cases with large x-axis coordinates correspond to PSFs with high BG and low signal and would likely be hard to detect under experimental conditions. These PSFs would therefore probably not be analyzed in localization microscopy. However, for single-particle tracking, this is not the case. When a fluorescently labeled object gradually bleaches away, one has high confidence in the presence of a dim object within a certain ROI due to the known trajectory from previous frames. Therefore, subtraction of the BG with BGnet can strongly increase the length of the whole trajectory, increasing the statistical strength of a diffusion analysis, for example. Furthermore, for brighter emitters which would be easily detected, BGnet remarkably still improves the localization precision by a factor of approximately two to ten (see the insets). For an additional analysis for the Tetra6 PSF with higher signal photon counts as typical for quantum dots or polystyrene fluorescent beads, see Fig. S10 and S11.

*BGnet strongly improves localization accuracy of single molecules for various BG complexities*
In the approach described above, the 100 PSF realizations were corrupted by different BG structures. The obtained position estimates were subsequently pooled to extract the spread of the localizations. While this method is intuitive, it does not report on the effect of an individual BG structure. To confirm that BG correction with BGnet improves the performance at the level of an individual localization event, we first developed a metric to quantify the complexity of the BG (termed "BG complexity") in a given PSF image. First, we calculated the spatial Fourier transform (FT) of the sBG alone. Additionally, we calculated the FT of a constant BG with the same average photon count per pixel and Poisson noise. Then, we subtracted the FT of the constant BG from the FT of the sBG to remove the dominant lowest spatial frequency. Next, we calculated the integrated weighted radial distribution. The result was normalized by the signal-to-background ratio (SBR; see Fig. S12 for details), yielding the BG complexity metric for the considered sBG, which is larger for BG with higher spatial frequencies or lower SBR. For each localization event, we calculated the Euclidian distance from the known true position (i.e., the accuracy) and plotted it against the respective BG complexity as depicted in Fig. 4.

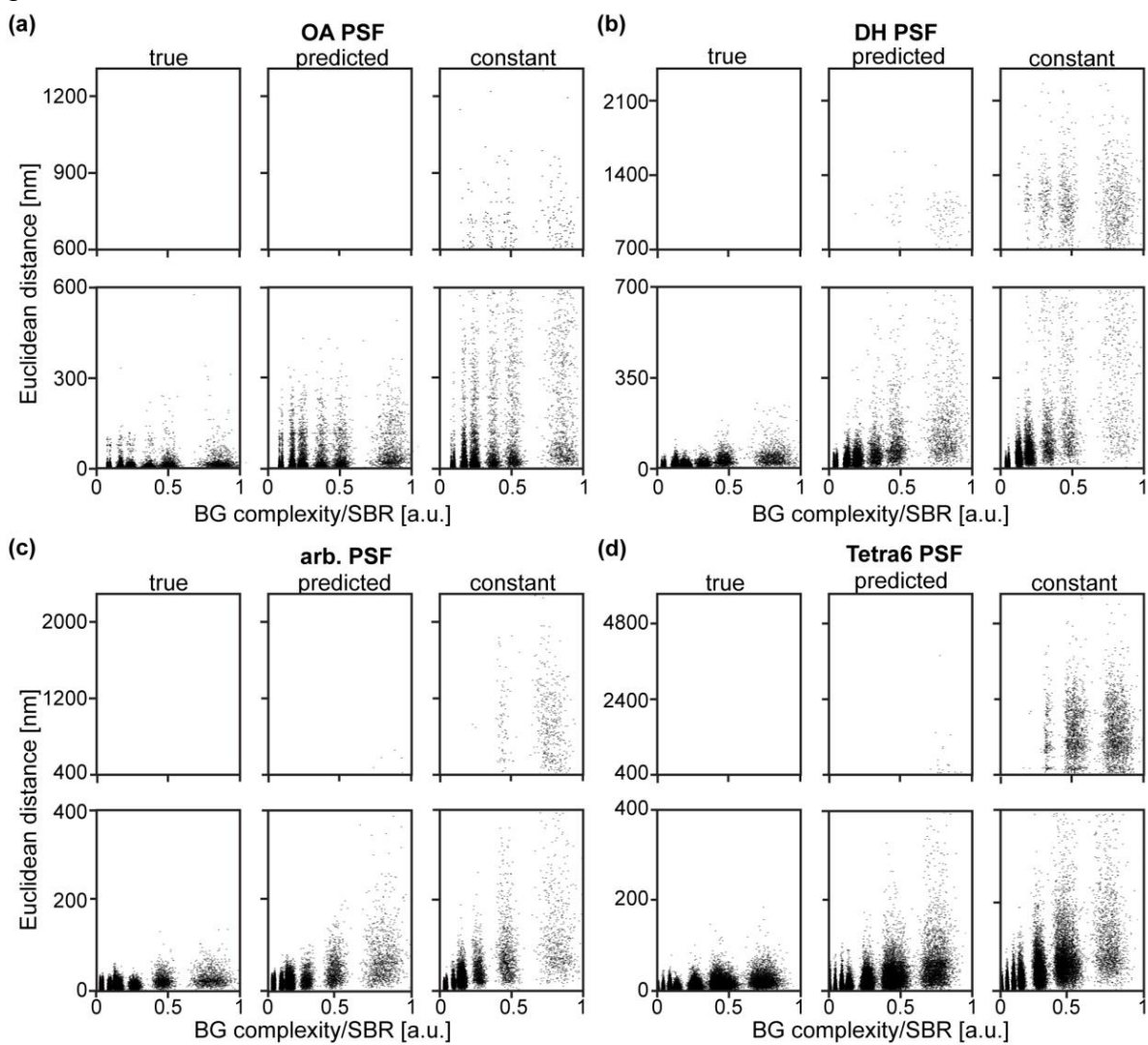

**Fig. 4:** Relationship between localization accuracy and BG complexity. To account for the influence of the SBR, which trivially has an effect on the localization accuracy, the BG

complexity metric was normalized by the SBR. For increasing BG complexity, the accuracy of the localization decreases, but in all cases, the BGnet outperforms constant BG subtraction. (a) to (d) OA PSF, DH PSF, arbitrary PSF, and Tetra6 PSF, respectively. True: BG correction with true BG; predicted: BG correction with the prediction from BGnet; constant: constant BG estimate. Note that the "streaks" visible arise from the discrete SBRs considered. Also note different y-axis scaling of top and bottom panels for each PSF.

This analysis confirms that BG correction with BGnet improves the accuracy of each single localization event. As is clearly visible, the differences between the estimated and the true positions are significantly smaller when the BG is corrected with BGnet compared to correction with a constant BG. This is true for all four analyzed PSF shapes. As one would expect, the accuracy decreases when the normalized BG complexity increases, regardless of the BG correction method (see bottom panels for each PSF – the scatter clouds rising from the x-axis). However, when the predicted BG from BGnet is used, this trend is clearly dampened. Thus, BG correction with BGnet performs much closer to the ideal case, i.e. BG correction with the true BG. Additionally, the number of significant outliers is strongly reduced compared to BG correction with constant BG (see top panels for each PSF). In an experimental setting, for example in localization microscopy, this is of high relevance as gross mislocalizations deteriorate image quality twofold: First, the number of spurious localizations in the reconstruction increases, and second, the localizations no longer report on the structure to be imaged, reducing the spatial resolution.

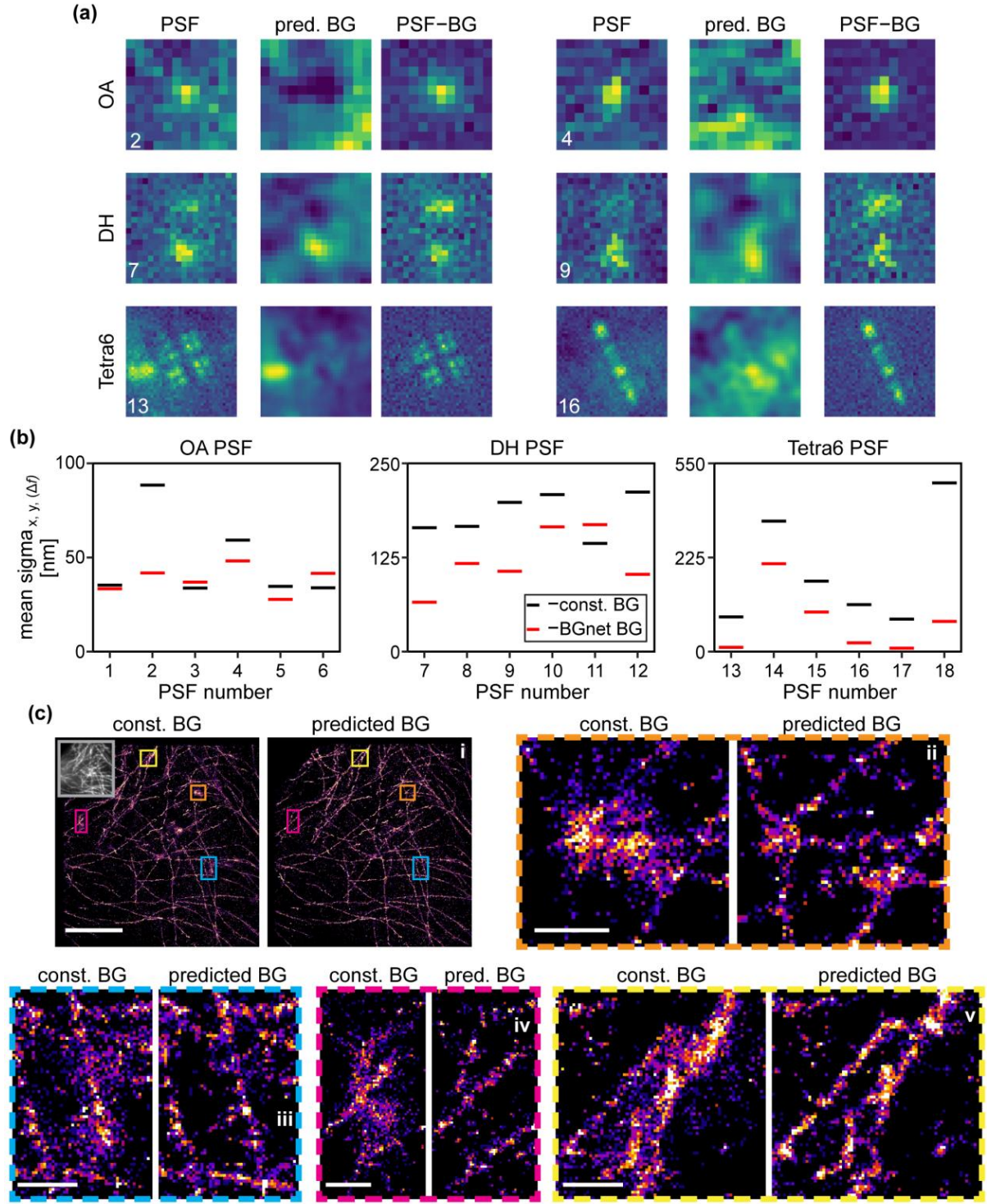

**Fig. 5:** Performance of BGnet on experimental data. (a) Two representative frames for three imaged PSF shapes, predicted BG, and corresponding BG-corrected PSFs. Contrast settings are not equal between the images. Background complexity values were 0.57 and 1.34 (OA PSF 2 and 4), 0.23 and 0.10 (DH PSF 7 and 9), and 0.34 and 0.35 (Tetra6 PSF 13 and 16), where the scaling is the same as in Figure 4. (b) Mean standard deviations of x, y, and $\Delta f$ position estimates over 1000 frames for six experimental realizations of each PSF shape, either

assuming constant BG (black) or using BG correction with BGnet (red). Note that the position estimates are shifted to the origin to facilitate comparison. For the OA PSF, only the x and y position estimates are considered. Note that the large spread of standard deviations arises from varying SBRs as well as different BG structures. (c) Super-resolution reconstructions of microtubules in fixed BSC-01 cells using the OA PSF and BG correction with a constant BG estimate or with BGnet. Four magnified regions are shown. Contrast settings are equal for each compared region. Scale bar = 5 µm for the image depicting the entire field of view (i) and 500 nm for zoom-ins (ii, iii, iv, and v). The inset subpanel i depicts the corresponding diffraction-limited image. Reconstructions are shown as 2D histograms with 23.4 nm bin width.

*BGnet enhances localization precision and image quality for experimental datasets*
Finally, we verify the performance of BGnet on experimental data. For this, we first imaged 100 nm fluorescent polystyrene beads in water that were attached to a glass coverslide using either no phase mask (OA), the double-helix, or the Tetra6 phase mask. sBG was introduced with a continuously moving white light source that illuminated the sample non-homogenously during data acquisition. Also, a large number of beads was not attached to the glass, but diffused freely in solution. Their emission contributed to the structured BG as well. For each PSF shape, we imaged different immobile beads for 1000 frames which were positioned at different regions of the field of view and exhibited different SBRs and BG structures. Then, we performed MLE fitting, either assuming a constant BG or performing BG correction with BGnet.

Fig. 5(a) shows representative frames from the obtained stacks for the three PSFs, the corresponding estimated BGs using BGnet, and the resulting BG-corrected PSFs. The results for BGnet are striking. For example, a part of a PSF caused by a diffusing bead is visible for the Tetra6 PSF 13 at the left edge, which is correctly identified by BGnet. Also, sBG with lower spatial frequency, visible from "humps" in the images, is accurately removed, leading to more pronounced PSF images for all three investigated PSFs. We also extracted the BG complexity metric using the same approach as for the simulated data and also scaled it identically to provide the same arbitrary units as in Fig. 4. For the ROIs shown in Fig. 5(a), the values range from 0.1 to 1.34 (see figure caption). Compiling the scaled BG complexities for all frames of all beads yielded the histograms shown in Figure S13. Importantly, some values are larger than 1, which was the highest value we realized when training BGnet. Nevertheless, BGnet still performed well. This indicates that our approach is robust and does not sharply decrease in performance when the boundary of the training parameter space is exceeded.

The visual impression translates to significantly improved localization precisions when performing MLE fitting. Fig. 5(b) shows the standard deviation of the position estimates, averaged over x, y and $\Delta f$ ($\Delta f$ only for the DH and Tetra6 PSF) for six cases for each PSF (see Fig. S14 for example scatter plots). The localization precision is evidently increased by BG correction with BGnet. Only very rarely, BGnet performs worse than when constant BG is assumed (PSF 3, PSF 6, and PSF 11). However, in these cases, BGnet also does not strongly reduce the localization precision. Therefore, in the worst case, BG correction with BGnet performs comparable to constant BG subtraction, but will, in the majority of cases, greatly improve the localization precision.

While BGnet improves localization precision in a proof-of-concept scenario, a further relevant assessment is to test its capability in a commonly encountered experimental setting. To this end, we investigated how BG correction with BGnet performs in localization-based super-resolution microscopy of a biological structure. We labeled microtubules in fixed BSC-01 cells via immunostaining, using AlexaFluor 647 as a fluorescent dye. Then, we acquired STORM super-resolution microscopy data and localized the detected single molecules. Also, we acquired a sBG image by illuminating an empty well with a LED white light source. We added this sBG image to each frame of the single-molecule localization data to introduce a strong sBG and thus to perform an assessment of BGnet under truly challenging conditions (see Fig. S15 for the sBG image and a representative frame). In the localization step, we either corrected for BG assuming a constant BG or using the estimate from BGnet (see Materials and Methods). The result is depicted in Fig. 5(c). The assumption of constant BG leads to severe artifacts in the reconstructions, evident from spurious localizations, non-structured regions, and loss of finer details. BG correction with BGnet, in contrast, yields excellent reconstructions of the microtubules (compare magnifications in the subpanels). Thus, we have successfully demonstrated the capability of BGnet to improve the image quality of super-resolution reconstructions, a result that can be readily transferred to other flavors of single-molecule experiments.

*Conclusion*
In summary, we have developed a robust and easy to implement method to rapidly correct PSF images for sBG. We demonstrate that this approach significantly improves emitter localization of OA, DH, and Tetra6 PSFs both for accurate PSF simulations and for experimental data. BGnet is not restricted to any specific assumptions about the BG characteristics. The method works because the PSF model is known, and it can be obtained accurately using known techniques. We hope that our method will improve PSF analysis for a wide range of powerful state-of-the-art techniques such as single-molecule localization microscopy,(36-38) single-molecule and single-particle tracking,(39) aberration correction with adaptive optics,(40) or deep-tissue imaging, where sBG is an especially prominent issue as recently highlighted by a noteworthy study.(41) Furthermore, we are confident that our workflow can be readily generalized to other flavors of microscopy and is not limited to just fluorescence microscopy, providing a broad range of scientific disciplines with a highly versatile resource.

**Materials and Methods**
*Cell culture:* BSC-01 cells were cultured in phenol red-free DMEM medium (Thermo Fisher, Waltham, MA), supplemented with 1mM sodium pyruvate (Thermo Fisher) and 10% FBS (Thermo Fisher), at 37 °C in a humidified 5 % $CO_2$ atmosphere. The cells were seeded into 8-well IBIDI chambered cover slides (ibidi, Gräfelfing, Germany) and used two days after seeding.

*Immunolabeling:* BSC-01 cells were washed with pre-warmed PBS +$Ca^{2+}$/$Mg^{2+}$ (Thermo Fisher) and pre-extracted with pre-warmed 0.2% saponin in CBS (10 mM MES, 138 mM NaCl, 3mM $MgCl_2$, 2 mM EGTA, 320 mM sucrose, all Sigma-Aldrich, St. Louis, MO) for 1 min. Then, cells were fixed with 3 % paraformaldehyde and 0.1 % glutaraldehyde (Sigma-Aldrich) in CBS for 15 min at room temperature (RT). Then, cells were reduced with 0.1 % $NaBH_4$ (Sigma-Aldrich) in PBS for 7 min at RT and rinsed 3 times for 3 min with PBS. Next, cells were blocked and permeabilized with 3 % bovine serum albumin (Sigma-Aldrich) (BSA) and 0.2 % Triton X-100 (Sigma-Aldrich) in PBS for 30 min at RT. Then, cells were incubated with the primary antibody (1:100 rabbit anti-alpha tubulin, ab18251, Abcam, Cambridge, UK) in 1 % BSA and 0.2 % Triton X-100 in PBS for 1 hour at RT, which was followed by 3 times 5-minute washes with 0.05 % Triton X-100 in PBS at RT. Then, cells were incubated with the secondary antibody (1:1000 donkey anti-rabbit AF647, ab150067, Abcam) in 1 % BSA and 0.2 % Triton X-1900 at RT. Finally, the cells were washed three times for 5 min with 0.05 % Triton X-100 in PBS at RT and post-fixed with 4 % paraformaldehyde for 10 min at RT. Finally, cells were washed three times 3 min each with PBS at RT and stored at 4 °C.

*Microscopy:* Cells were imaged on a custom epifluorescence microscope using a Nikon Diaphot 200 as core (Tokyo, Japan), equipped with an Andor Ixon DU-897 EMCCD camera (Belfast, UK), a high-N.A. oil-immersion objective (UPlanSapo 100×/1.40 N.A, Olympus, Tokyo, Japan), a motorized xy-stage (M26821LOJ, Physik Instrumente, Karlsruhe, Germany), and a xyz-pizeo stage (P-545.3C7, Physik Instrumente). Molecules were excited with a 642 nm 1W CW laser (MPB Communications Inc., Pointe-Claire, Canada). The emission was passed through a quadpass dichroic mirror (Di01-R405/488/561/635, Semrock, Rochester, NY) and filtered using a ZET642 notch filter (Chroma, Bellows Falls, VT) and a 670/90 bandpass filter (Chroma). For 3D imaging, DH (Double Helix Optics, Boulder, CO), and Tetra6 phase masks (described in reference (42)) were inserted into the 4f-system of the microscope as described previously.(43)

*MLE fitting algorithm:* In order to determine the position, signal photon counts, and background photon counts of single-emitter images, a maximum likelihood fitting algorithm was employed. Under the assumption of Poisson noise statistics, the objective function for maximum likelihood estimation is given by $\sum_i \mu_i(\theta) - n_i \ln(\mu_i(\theta))$, where $n_i$ is the photon count measured in pixel $i$ and $\mu_i(\theta)$ is the total photon count predicted in that pixel by a forward model of the point spread function for specific values of emitter parameters $\theta$ (position, signal photons, and background photons). Minimizing the objective function with respect to $\theta$ yields the maximum-likelihood parameter estimates $\hat{\theta}$.

*Super-resolution data acquisition and image reconstruction:* For super-resolution data acquisition, a reducing and oxygen scavenging buffer was used,(44) consisting of 40 mM cysteamine, 2 μL/mL catalase, 560 μg/mL glucose oxidase (all Sigma-Aldrich), 10% (w/v) glucose (BD Difco, Franklin Lakes, NJ), and 100 mM Tris-HCl (Thermo Fisher). The exposure time was 30 ms and the calibrated EM gain was 186. Single-molecule signals were detected with a standard local maximum intensity approach. Each single molecule signal was fitted to a 2D Gaussian, either without BG correction using BGnet or with BG correction using BGnet. In both cases, a constant offset was implemented for the fitting. If no BG correction with BGnet was applied, this translates to an estimated constant BG. For initial BG correction with BGnet, the offset was, expectedly, very close to zero. The position of the maximum of the Gaussian fit was stored as the localization of the single molecule. Drift correction was performed via cross correlation.

*Data and code supporting the findings of this manuscript are available from the corresponding author upon reasonable request.*

**Funding:** This work was supported in part by the National Institute of General Medical Sciences Grant No. R35GM118067. PNP is a Xu Family Foundation Stanford Interdisciplinary Graduate Fellow.

**Acknowledgments:** We thank Kayvon Pedram for stimulating discussions and Anna-Karin Gustavsson for cell culture.

Electronic Supplemental Information to
# BGnet: Accurate and rapid background estimation with deep neural nets

Leonhard Möckl, Anish R. Roy, Petar N. Petrov, and W.E. Moerner

## Table of contents

**Table S1:** OA PSF simulation parameters.

| Parameter | Value |
| --- | --- |
| image size | 12x12 px |
| pixel size | 117.2 nm |
| signal photons | 2,500 to 10,000/PSF |
| average background photons | 25 to 150/px |
| mean no-light counts | 101 |
| σ(no-light counts) | 4.15 |
| no-light counts noise model | Gaussian |
| photon noise model | Poisson |
| conversion gain | 26.93 counts/photoelectron |
| EM gain | 186 |
| emitter z-position range | 0 to 4 µm |
| z-step | 500 nm |
| focal position range | -0.5 to 0.5 µm |
| f-step | 50 nm |
| $\lambda_{emission}$ | 671 nm |
| dipole contributions in x, y, z | (1,1,1)/3 |
| $NA_{objective}$ | 1.4 |
| $n_{coverslip}$ | 1.518 |
| $n_{medium}$ | 1.33 |

**Table S2:** DH PSF simulation parameters.
Identical to OA PSF parameters except:

| Parameter | Value |
| --- | --- |
| image size | 20x20 px |
| signal photons | 2,500 to 10,000/PSF |
| average background photons | 25 to 150/px |
| emitter z-position range | 0 to 4 µm |
| z-step | 500 nm |
| focal position range | -1 to 1 µm |
| f-step | 100 nm |

**Table S3:** Arbitrary PSF simulation parameters.
Identical to OA PSF parameters except:

| Parameter | Value |
| --- | --- |
| image size | 20x20 px |
| signal photons | 5,000 to 25,000/PSF |
| average background photons | 25 to 150/px |
| emitter z-position range | 0 to 4 µm |
| z-step | 500 nm |
| focal position range | -1 to 1 µm |
| f-step | 100 nm |

**Table S4:** Tetra6 PSF simulation parameters.
Identical to OA PSF parameters except:

| Parameter | Value |
| --- | --- |
| image size | 40x40 px |
| signal photons | 10,000 to 30,000/PSF |
|  | 7.500 to 75,000/PSF (analysis 2) |
| average background photons | 25 to 150/px |
| emitter z-position range | 0 to 10 µm |
| z-step | 500 nm |
| focal position range | -2.5 to 2.5 µm |
| f-step | 50 nm |

**Table S5:** Training parameters.

| Parameter | Value |
|---|---|
| # of PSFs stacks for training | 18,000 (OA PSF)<br>18,000 (DH PSF)<br>75,000 (arbitrary PSF)<br>200,000 (Tetra6 PSF) |
| # of PSF stacks for validation | 3,500 (OA PSF)<br>3,500 (DH PSF)<br>3,500 (arbitrary PSF)<br>5,000 (Tetra6 PSF) |
| optimizer | Adam ($\beta_1=0.9$, $\beta_2=0.999$, $\varepsilon=1E-8$) |
| loss function | MSE |
| initial learning rate (LR) | 0.002 |
| $\varepsilon$ for LR decrease | 1E-7 |
| factor for LR decrease | 0.5 |
| patience | 3 epochs |
| minimal LR | 1E-6 |
| batch size | 64 |

**Table S6:** Simulation parameters for MLE analysis.

| Parameter | Value |
|---|---|
| signal photons | 2,500, 6250, 10,000/PSF (OA PSF)<br>2,500, 6250, 10,000/PSF (DH PSF)<br>5,000, 15,000, 25,000/PSF (arbitrary PSF)<br>10,000, 20,000, 30,000/PSF (Tetra6 PSF)<br>7,500, 42150, 75,000/PSF (Tetra6 PSF, analysis 2) |
| average background photons | 25, 87.5, 150/px (all PSFs) |
| emitter z-positions | 0, 2 µm (OA PSF)<br>0, 2 µm (DH)<br>0, 2 µm (arbitrary PSF)<br>0, 2, 4, 6, 8, 10 µm (Tetra6 PSF) |
| focal positions | -0.5, -0.25, 0, 0.25, 0.5 (OA PSF)<br>-1, -0.5, 0, 0.5, 1 µm (DH PSF)<br>-1, -0.5, 0, 0.5, 1 µm (arbitrary PSF)<br>-2.5, -1.25, 0, 1.25, 2.5 µm (Tetra6 PSF) |

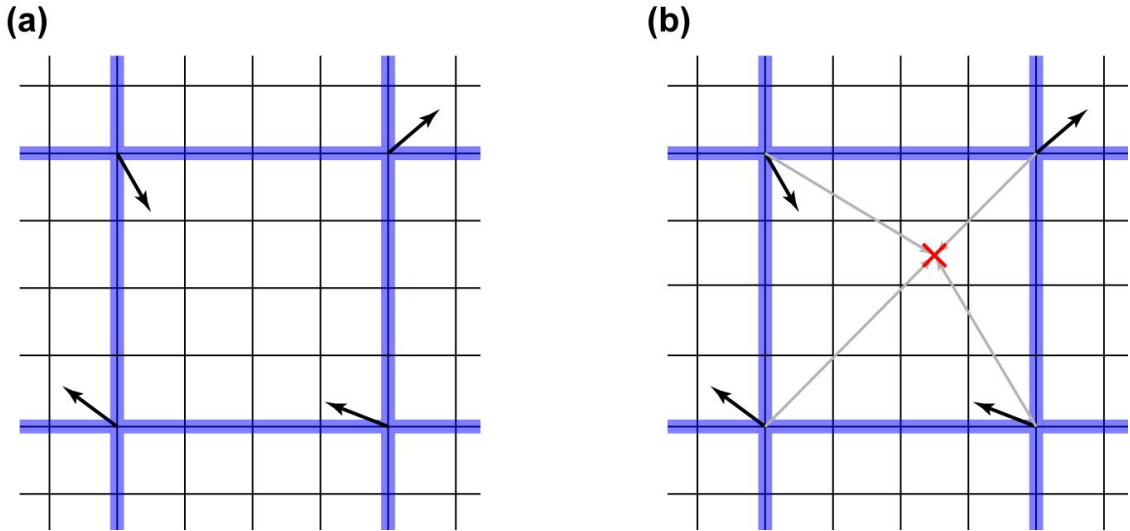

**Figure S1:** Principle of Perlin noise simulation. (a) The image pixels (black) are overlaid with a grid of superpixels (blue). The length of the superpixels (in this case four image pixels) determines the spatial frequency of the noise. To each corner of the superpixel grid, a gradient vector with random orientation is assigned (black vectors). (b) For each image pixel, vectors from the four closest corners to the pixel are calculated (red cross and grey vectors). Then, the dot product of these corner vectors with the four corresponding gradient vectors at the corners are calculated. Each of the resulting dot products is multiplied with the fade function $f(u) = 6u^5 - 15\,u^4 + 10u^3$, where $u$ is the position of the image pixel in the coordinate system of the superpixel grid (that is, $u$ is always between zero and one). The fade function has zero first and second derivatives at $u = 0$ and $u = 1$, which ensures continuous noise at the edges of the superpixel grid. The attenuated dot products are then summed, yielding the noise at the specific image pixel. This process is repeated for all image pixels. By combining weighted Perlin noise images with different superpixel widths, noise with defined contributions from different spatial frequencies can be realized.

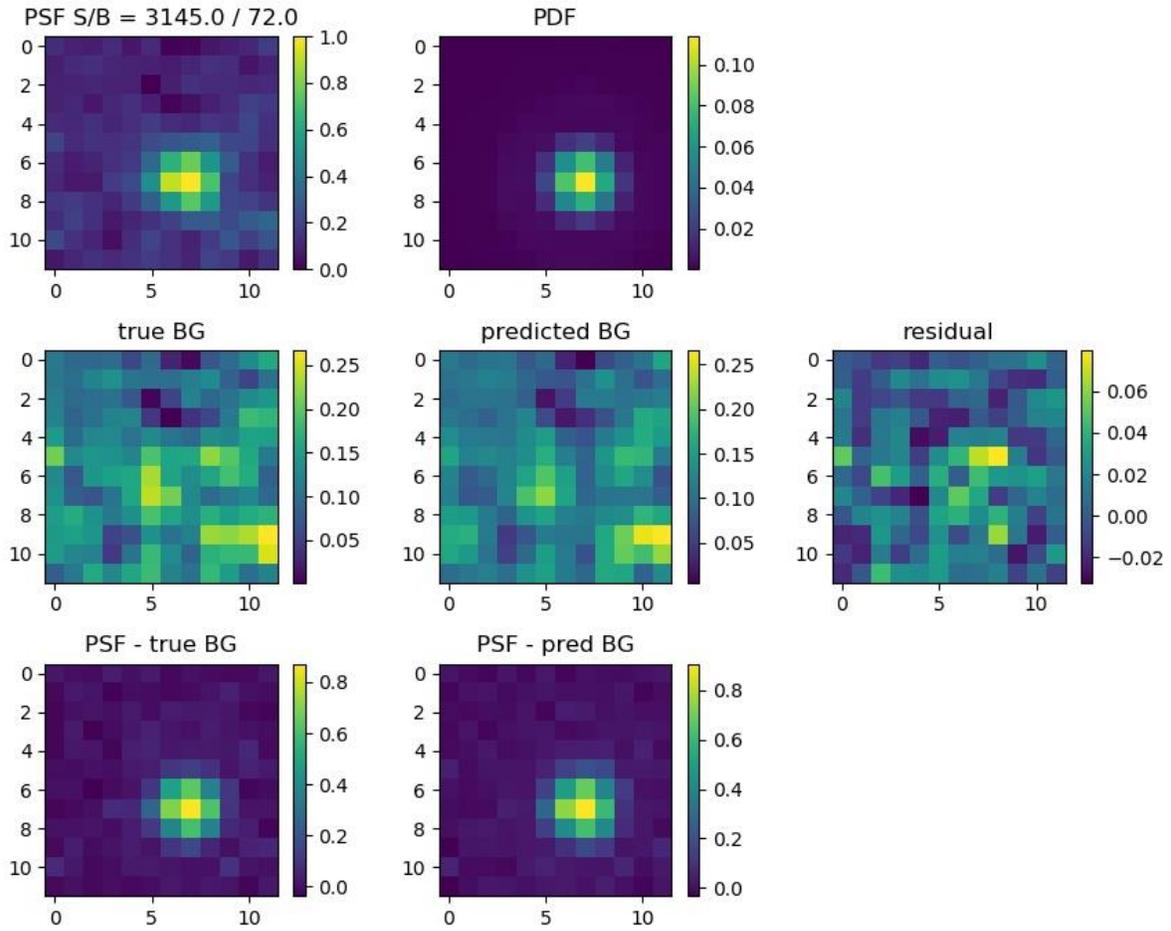

**Figure S2:** Representative examples for accurate BG: OA PSF. S/B indicates signal photons and average BG photons per pixel.

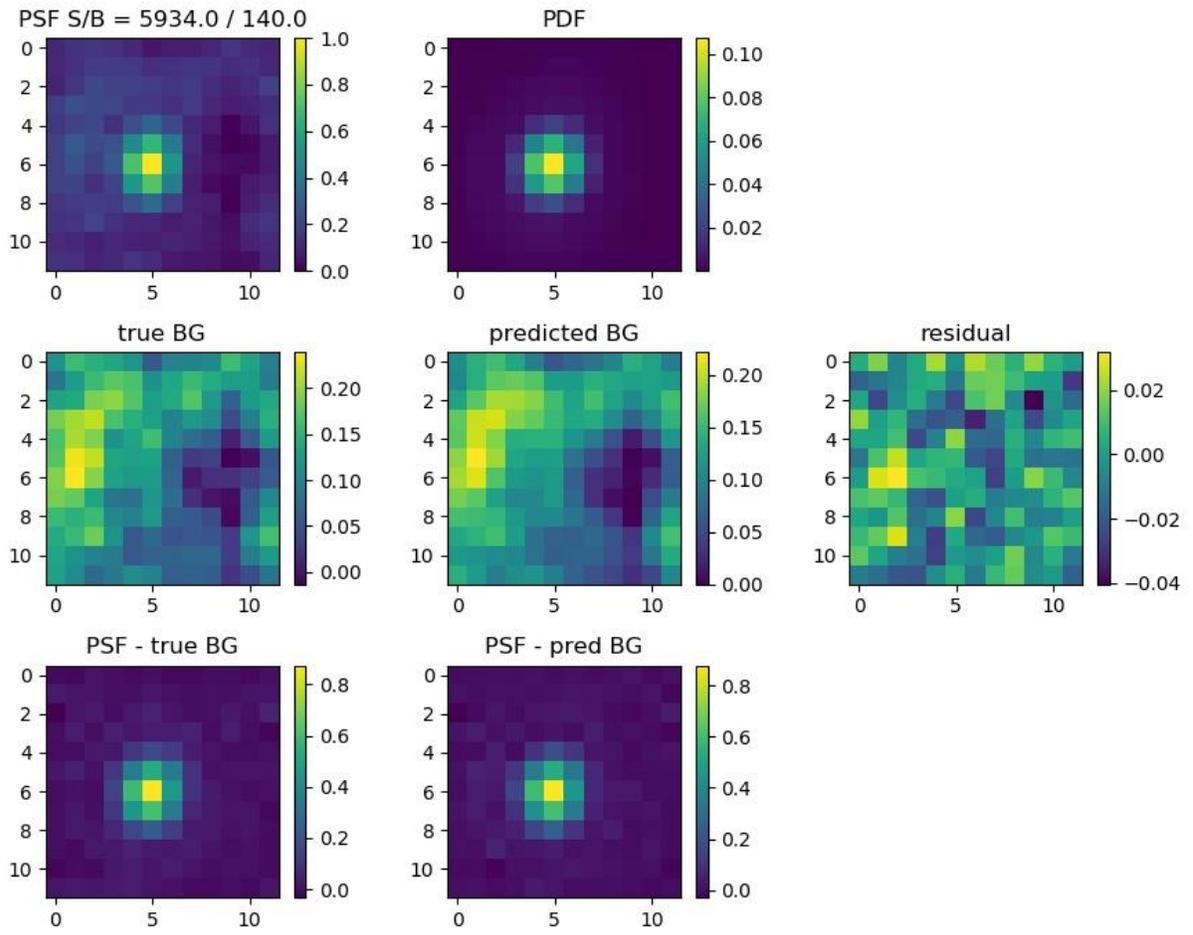

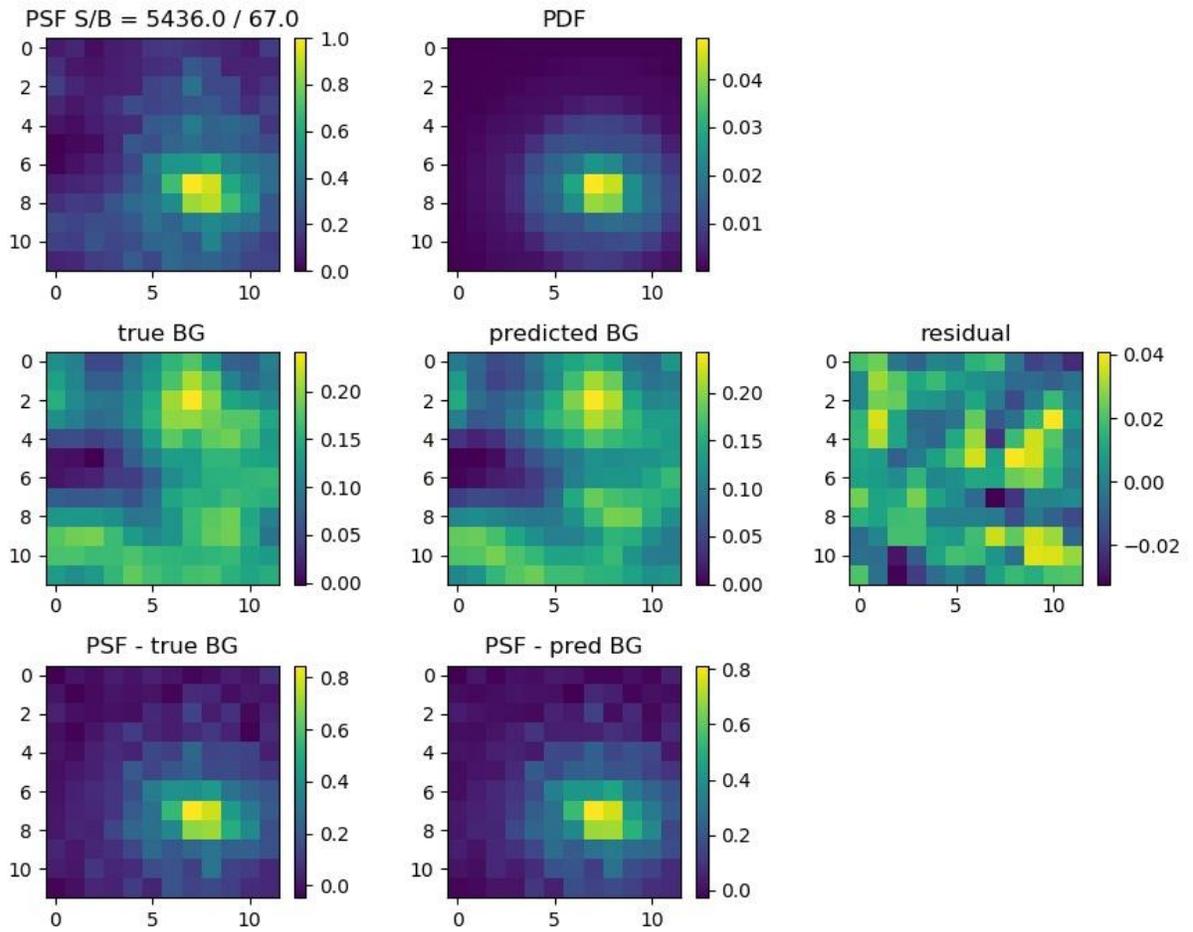

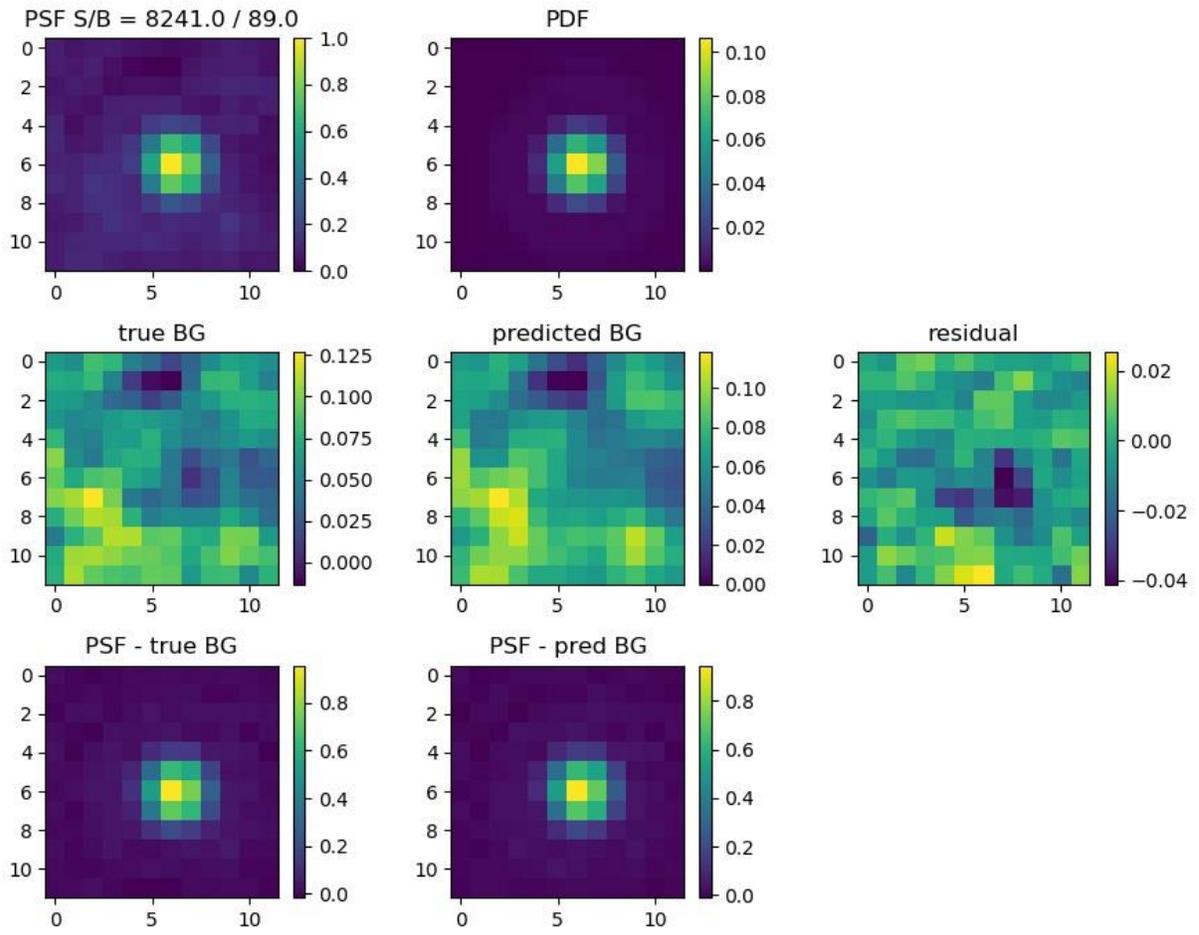

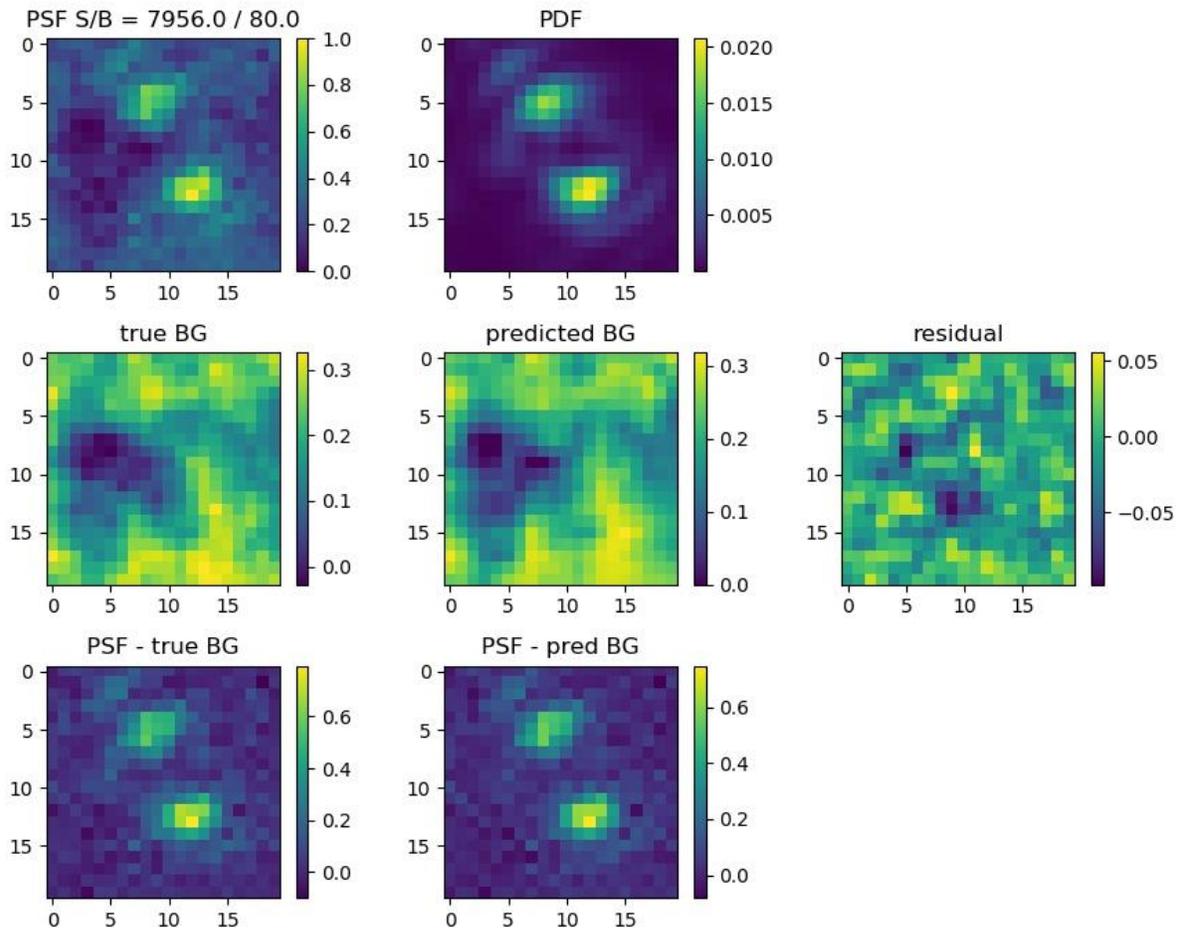

**Figure S3:** Representative examples for accurate BG: DH PSF. S/B indicates signal photons and average BG photons per pixel.

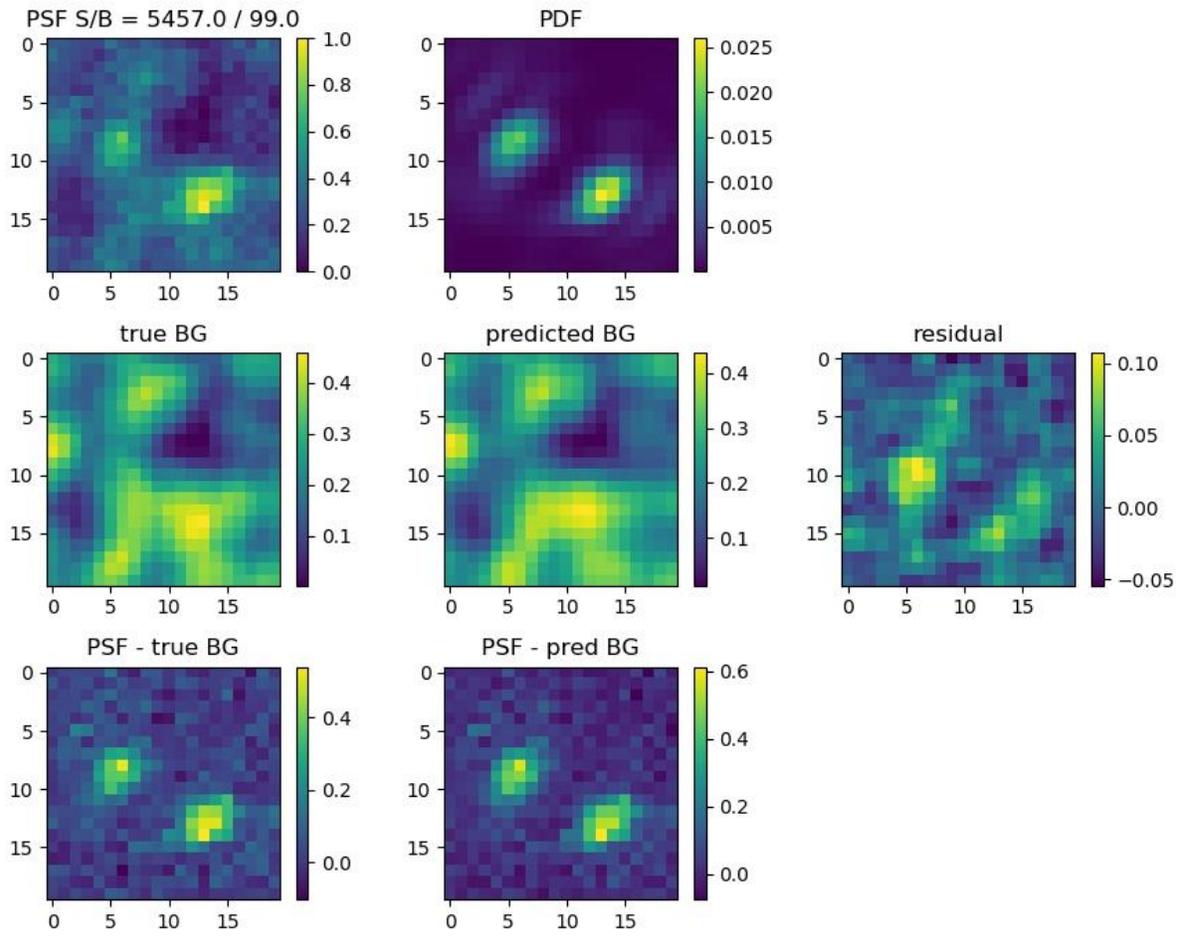

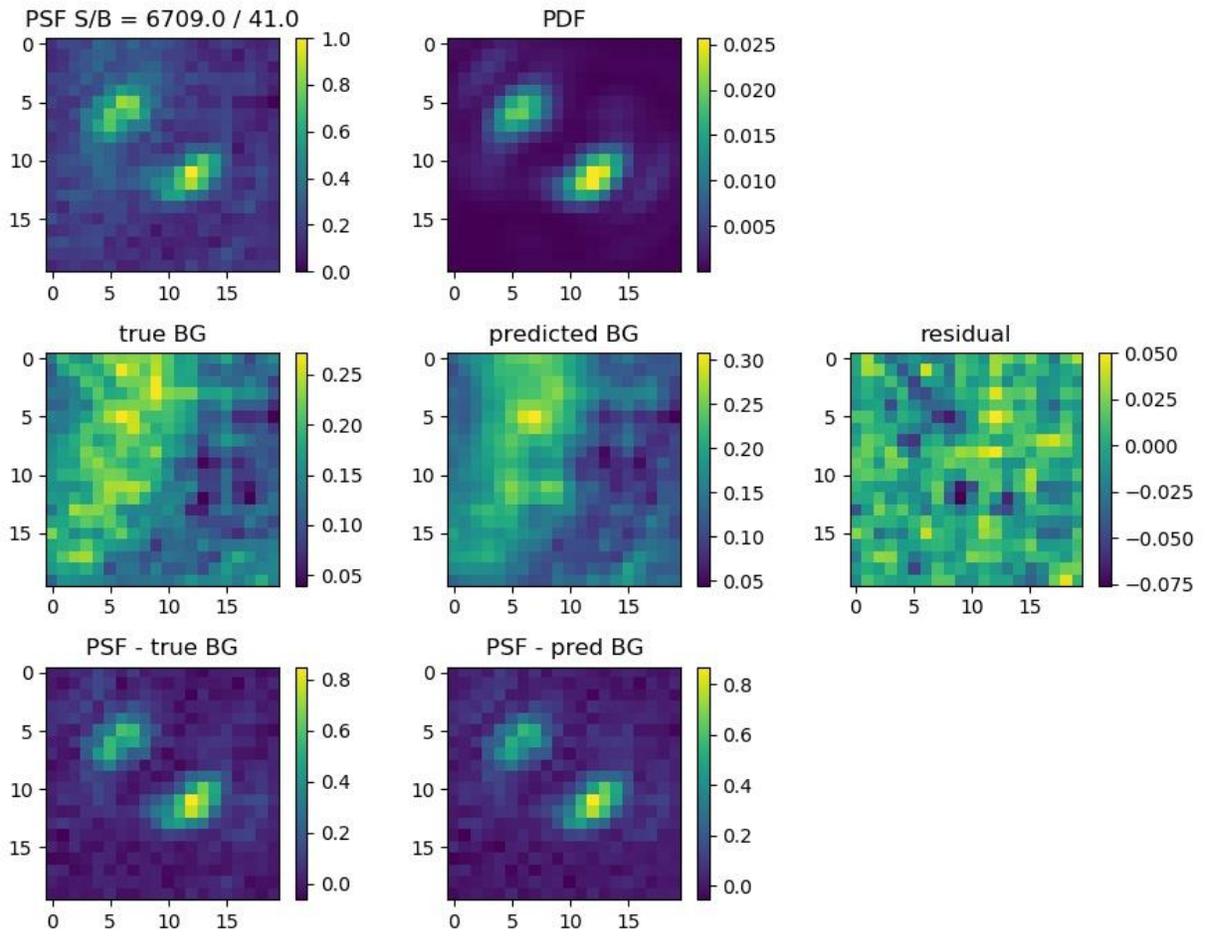

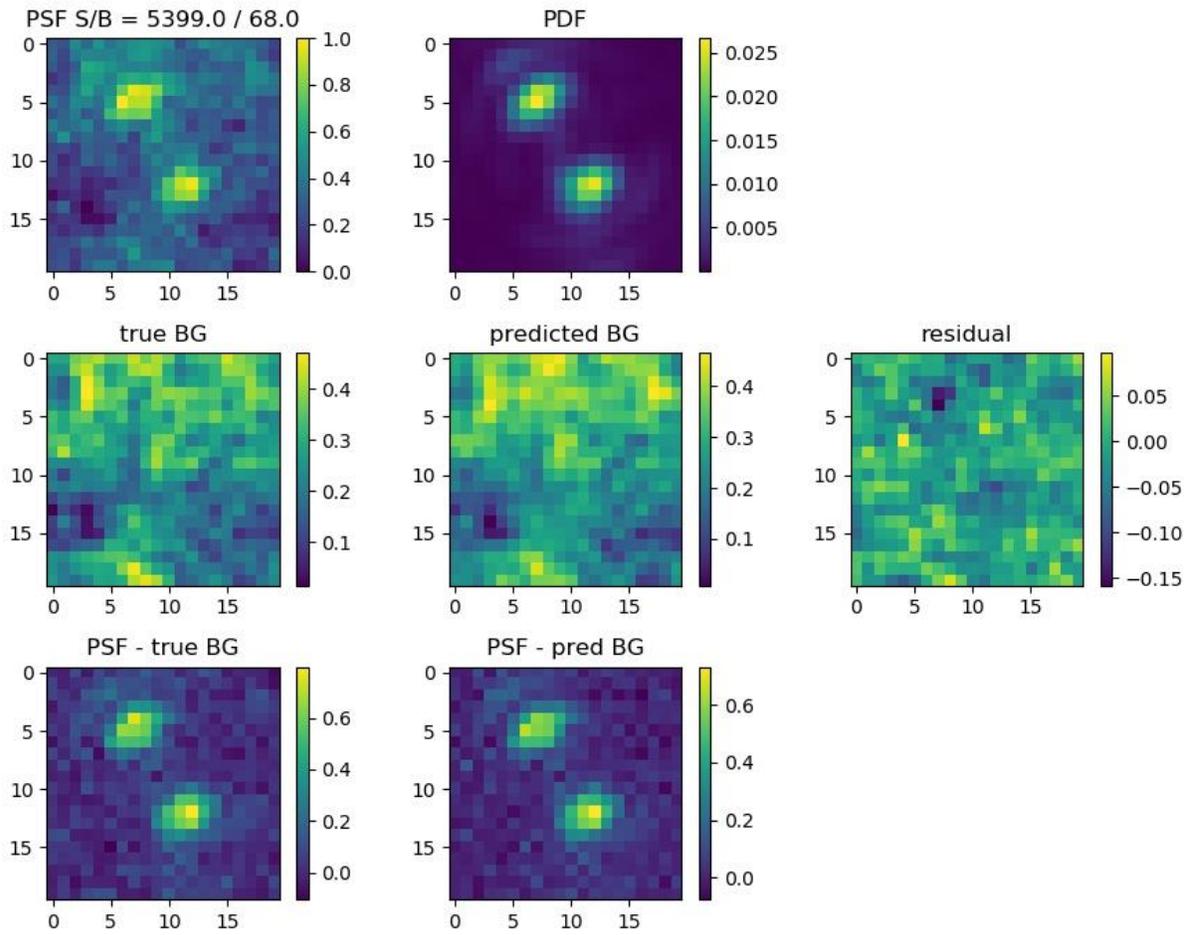

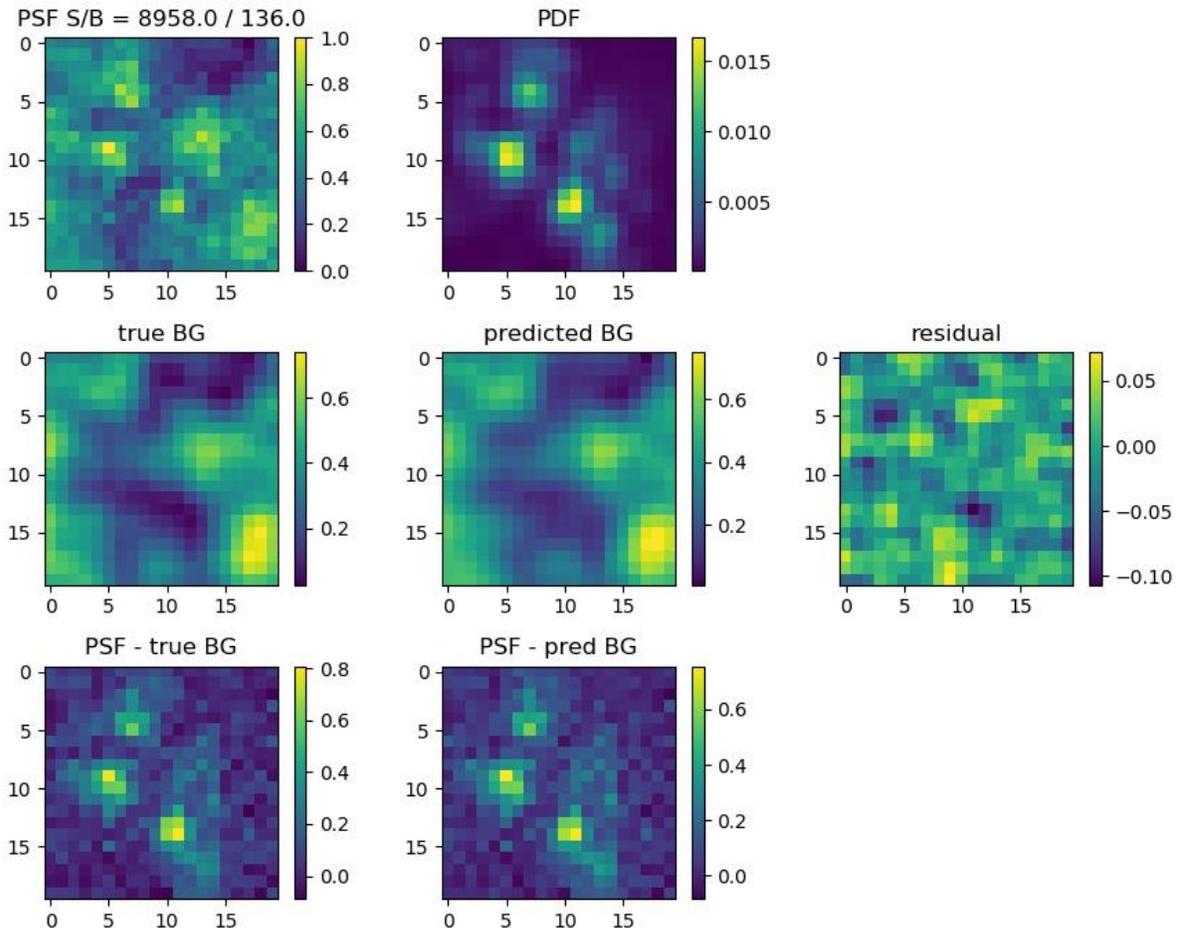

**Figure S4:** Representative examples for accurate BG: Arbitrary PSF. S/B indicates signal photons and average BG photons per pixel.

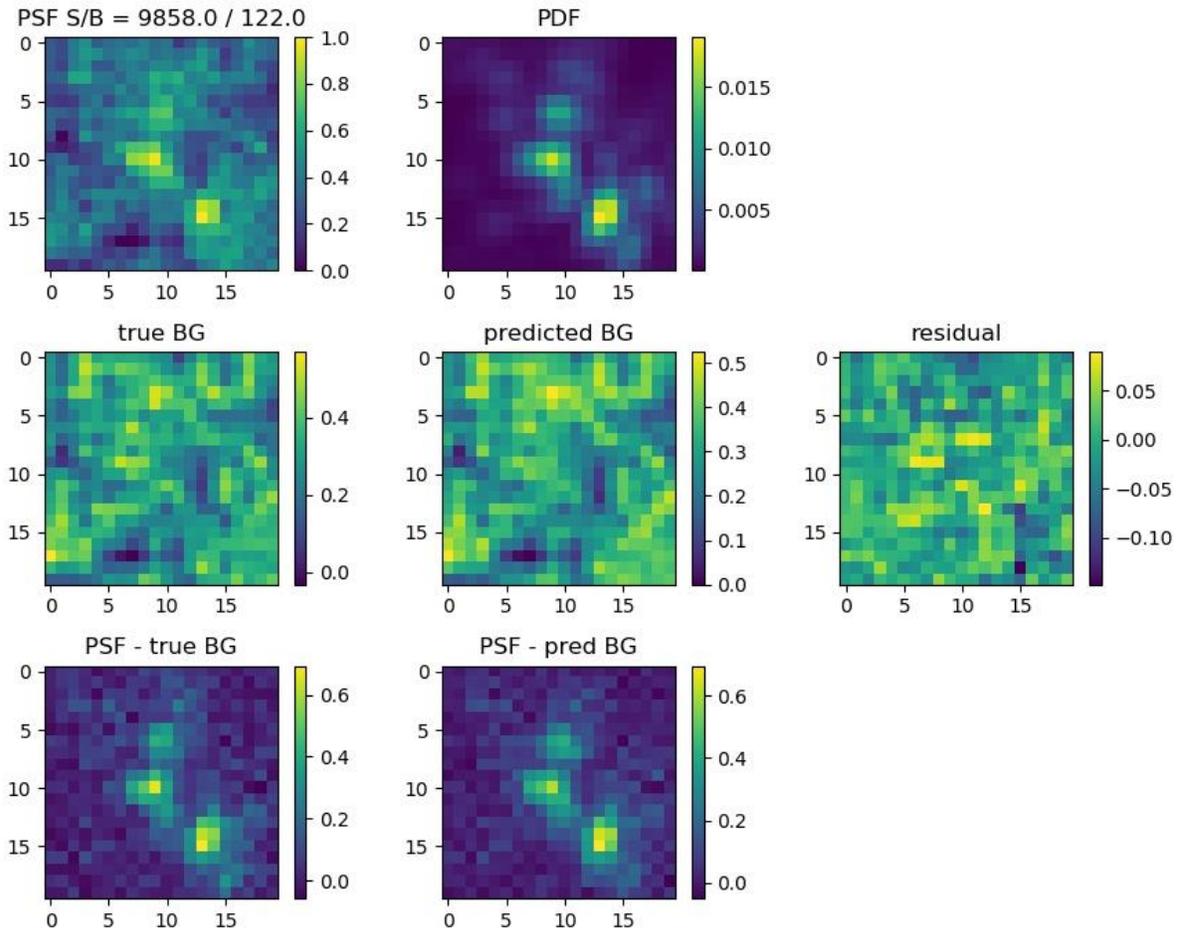

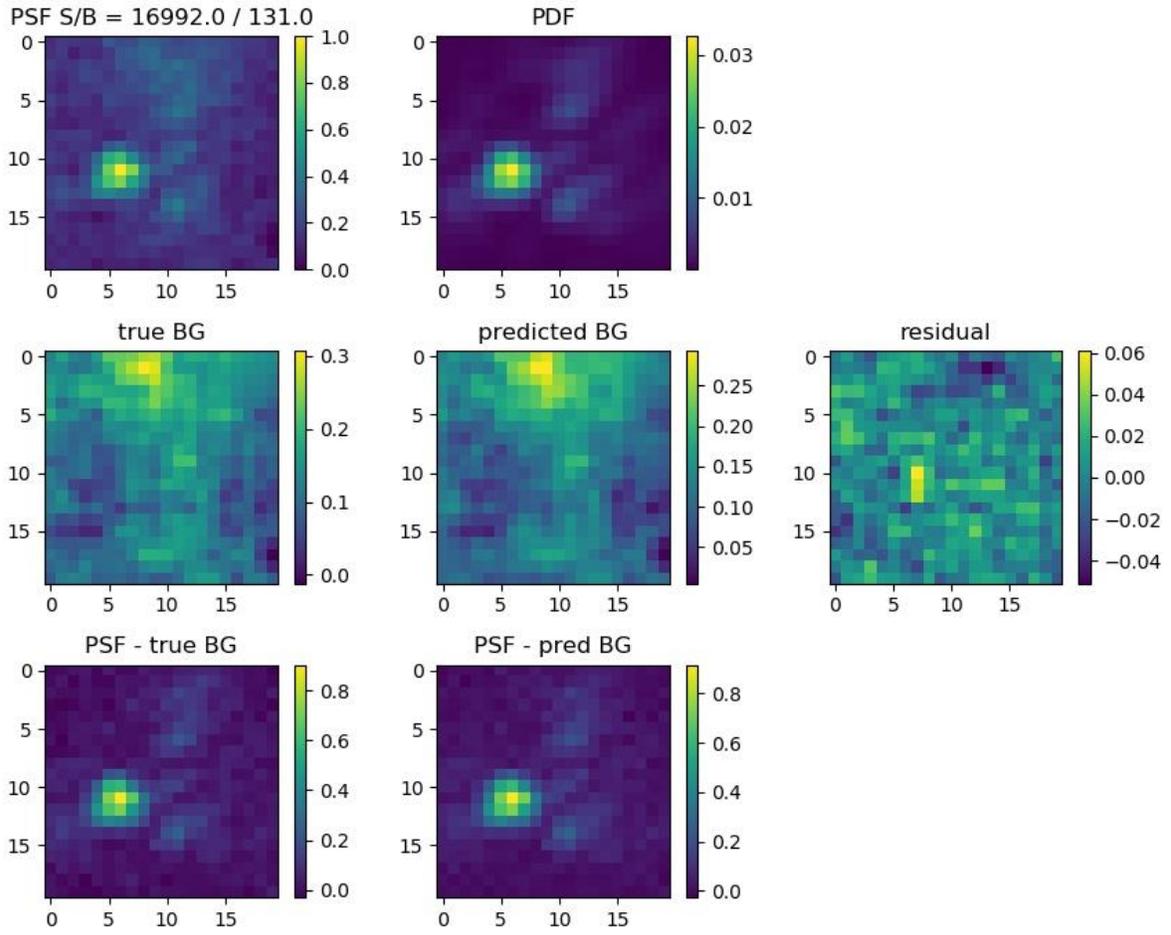

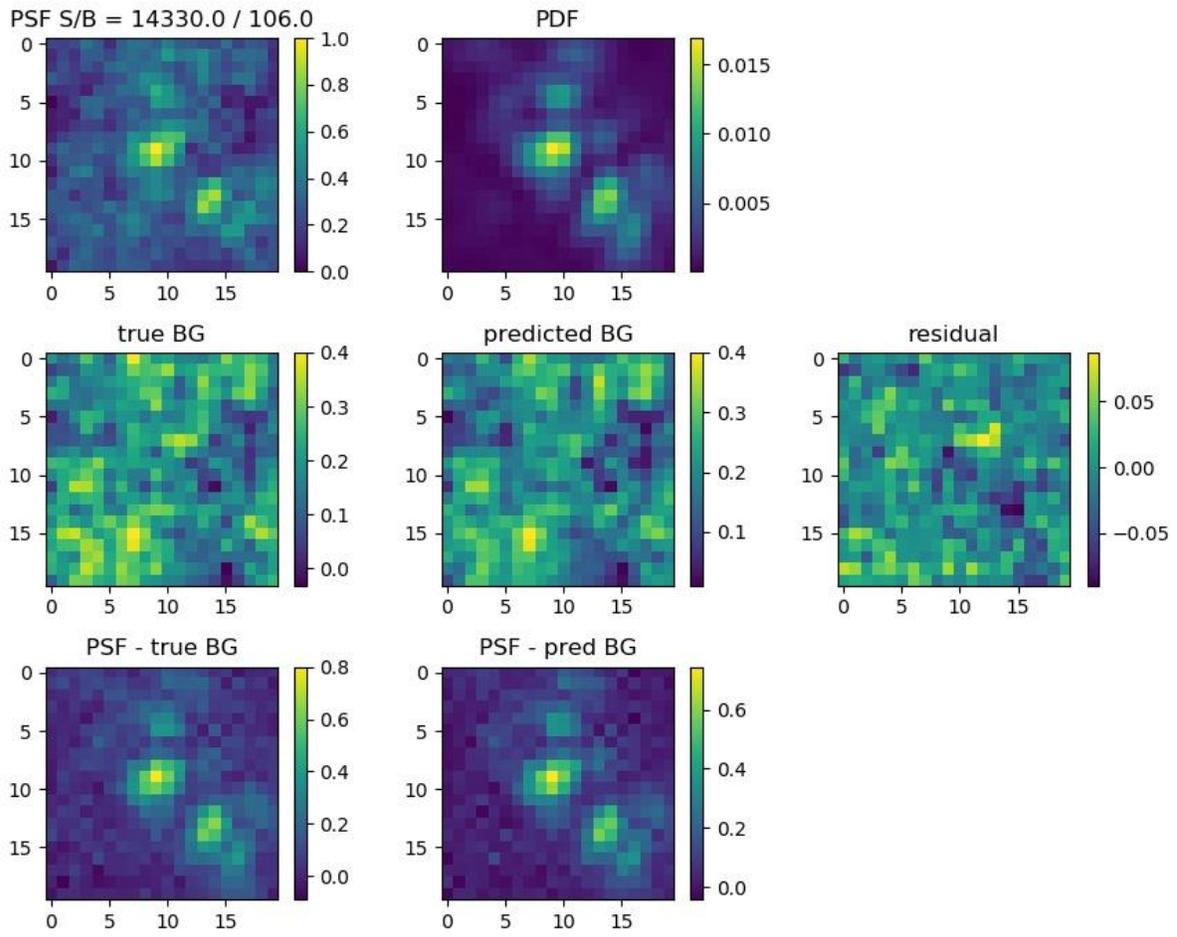

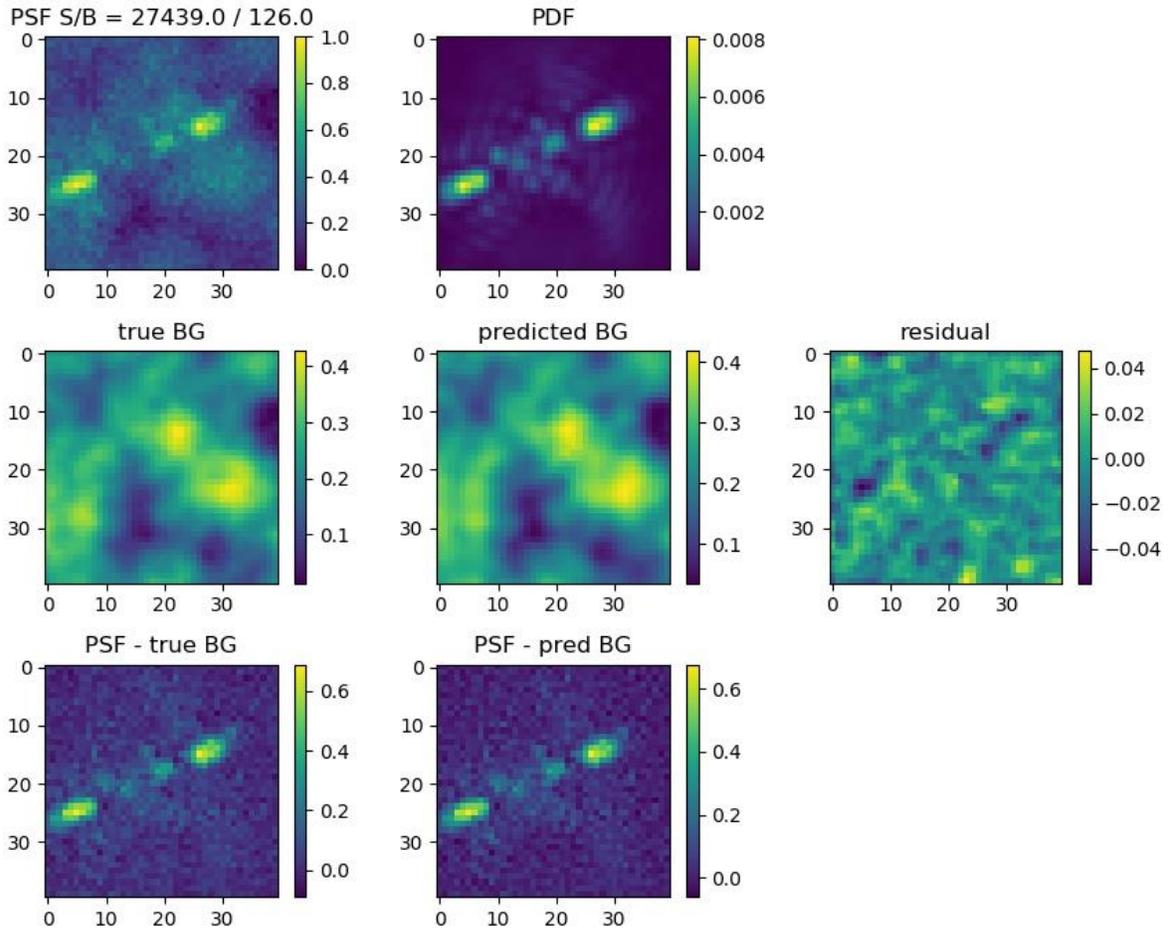

**Figure S5:** Representative examples for accurate BG: Tetra6 PSF. S/B indicates signal photons and average BG photons per pixel.

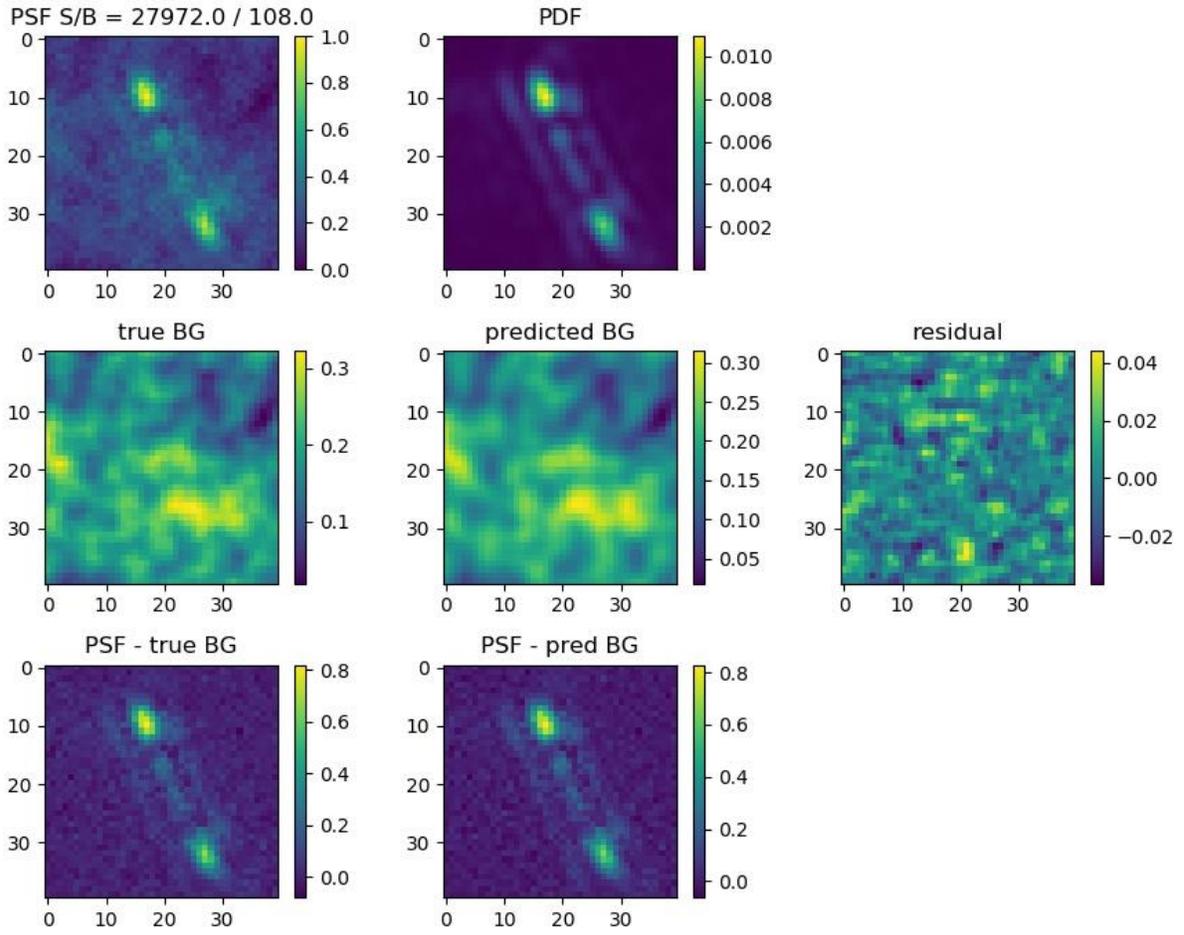

(continuation of Figure S5)

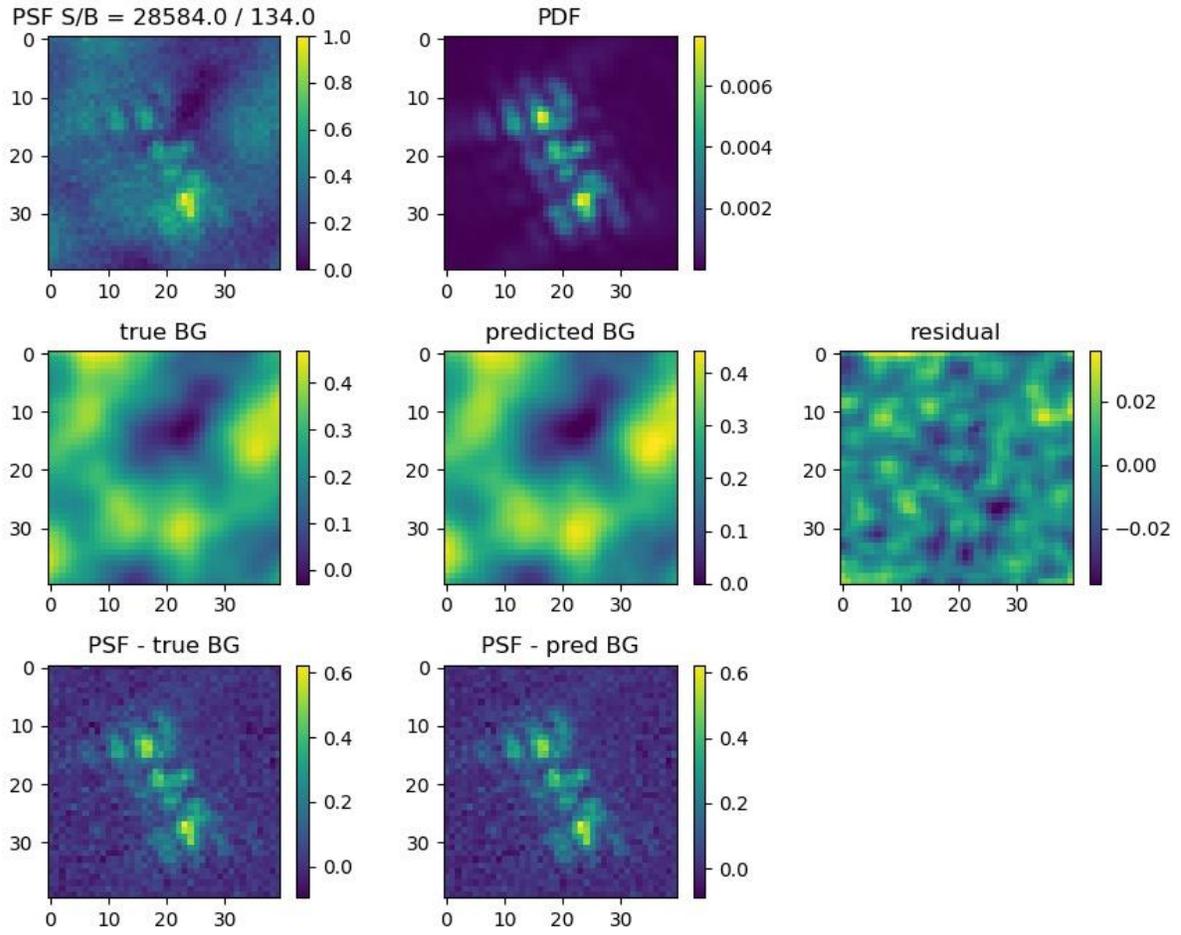

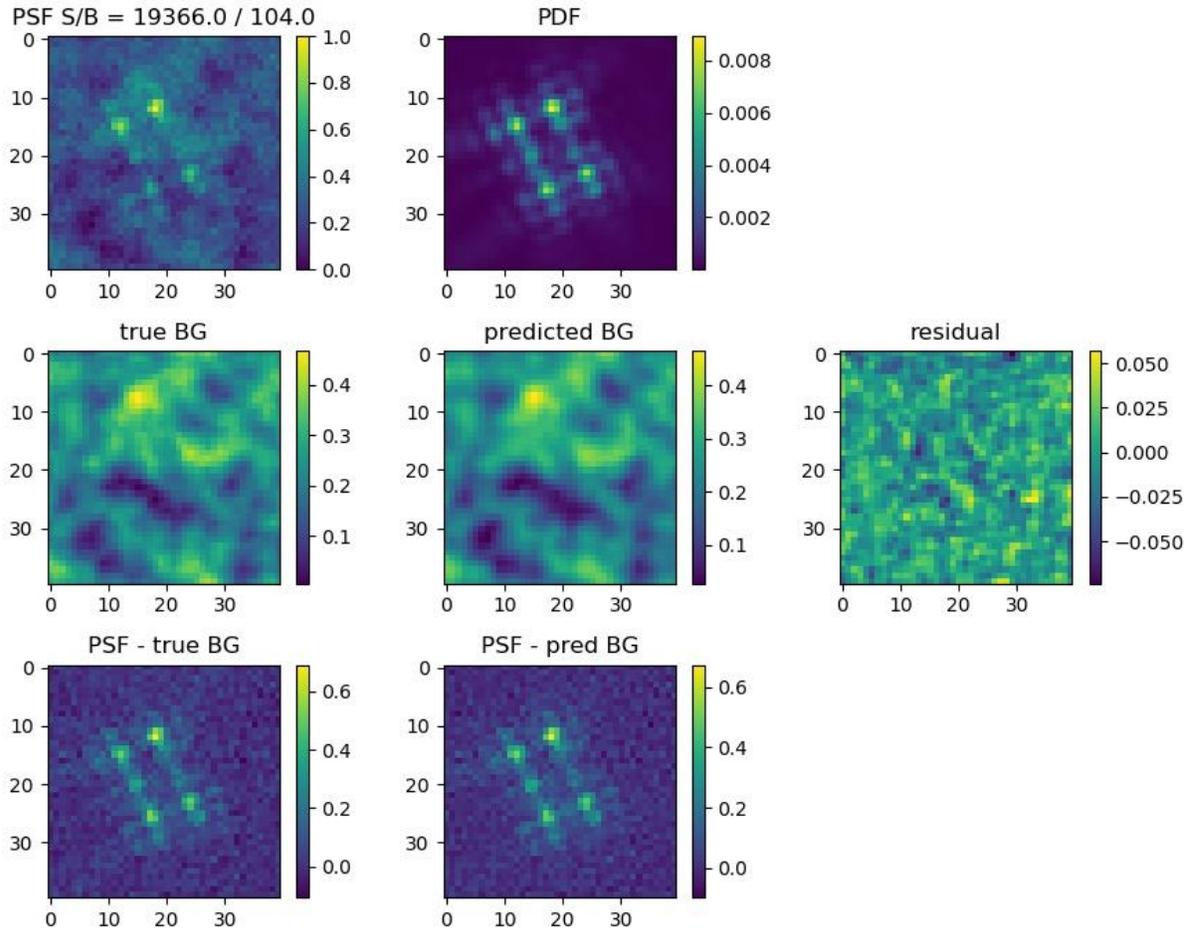

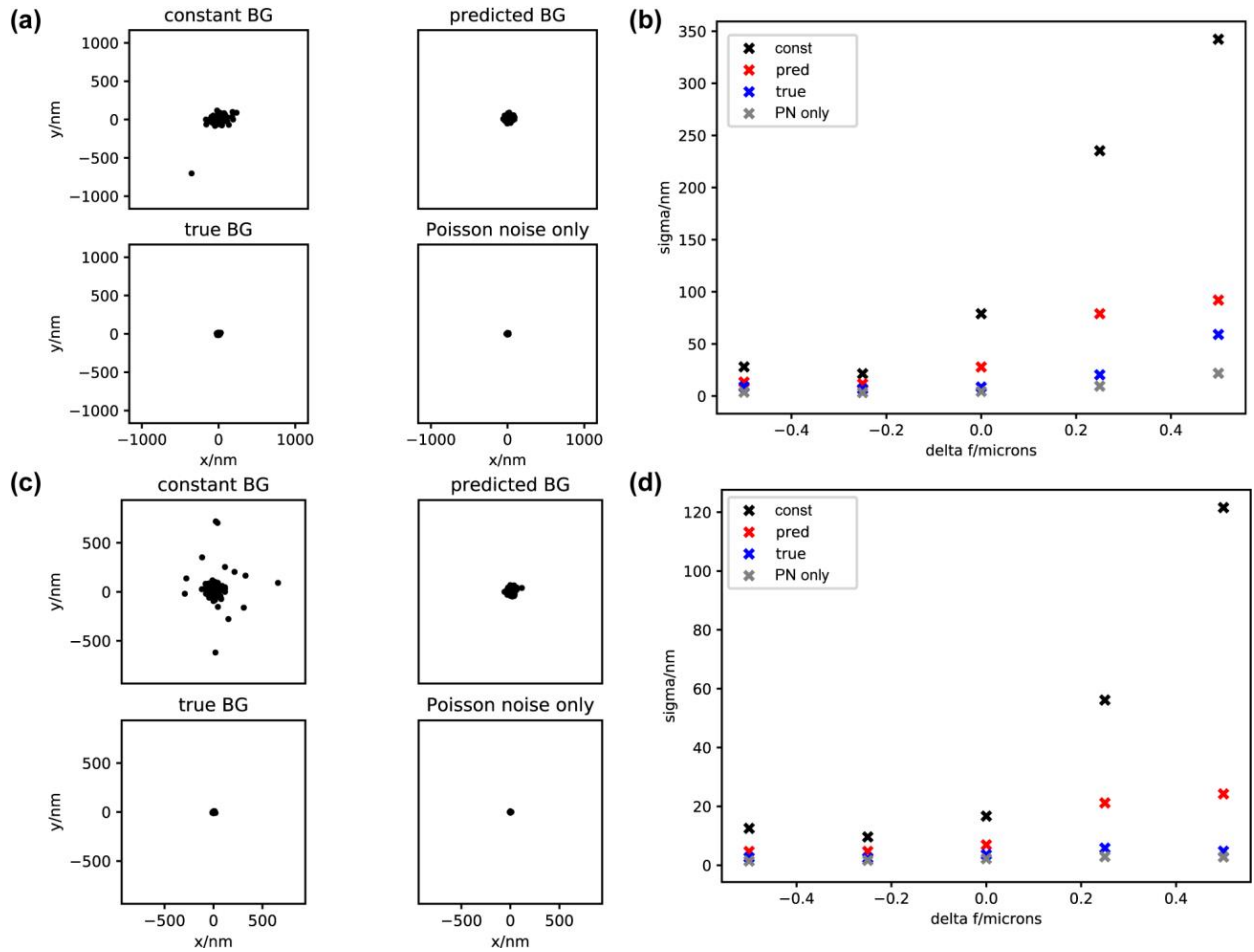

**Figure S6:** Example Scatter plots and corresponding standard deviations for MLE analysis of the OA PSF. (a) and (b) 2,500 signal photons, 87.5 average BG photons. Emitter at 0 µm, focal position for scatter plot at 0 µm. (c), (d) 10,000 signal photons, 150 average BG photons. Emitter at 2 µm, focal position for scatter plot at 0.5 µm.

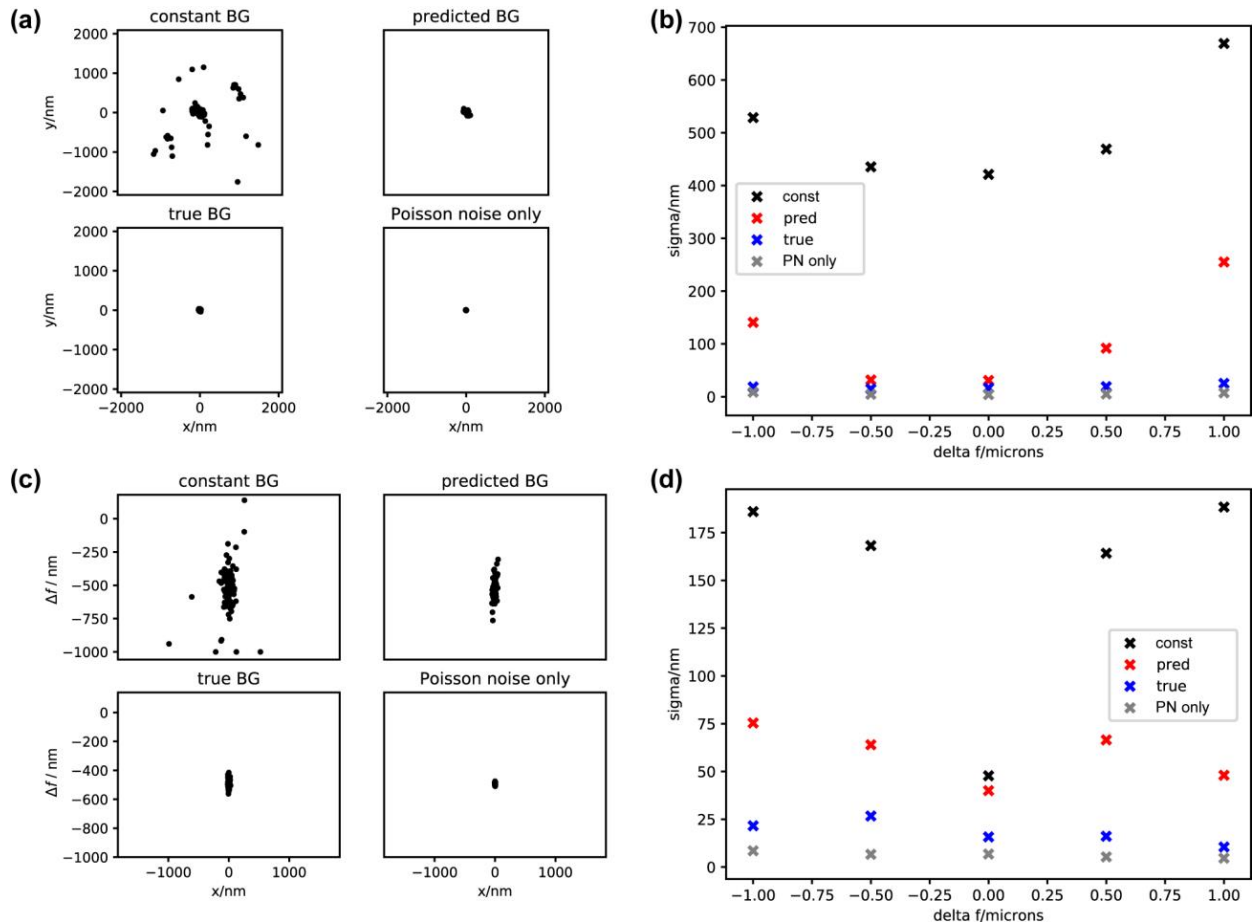

**Figure S7:** Example Scatter plots and corresponding standard deviations for MLE analysis of the DH PSF. (a) and (b) 2,500 signal photons, 87.5 average BG photons. Emitter at 0 µm, focal position for scatter plot at -0.5 µm. (c), (d) 6,250 signal photons, 150 average BG photons. Emitter at 2 µm, focal position for scatter plot at -0.5 µm.

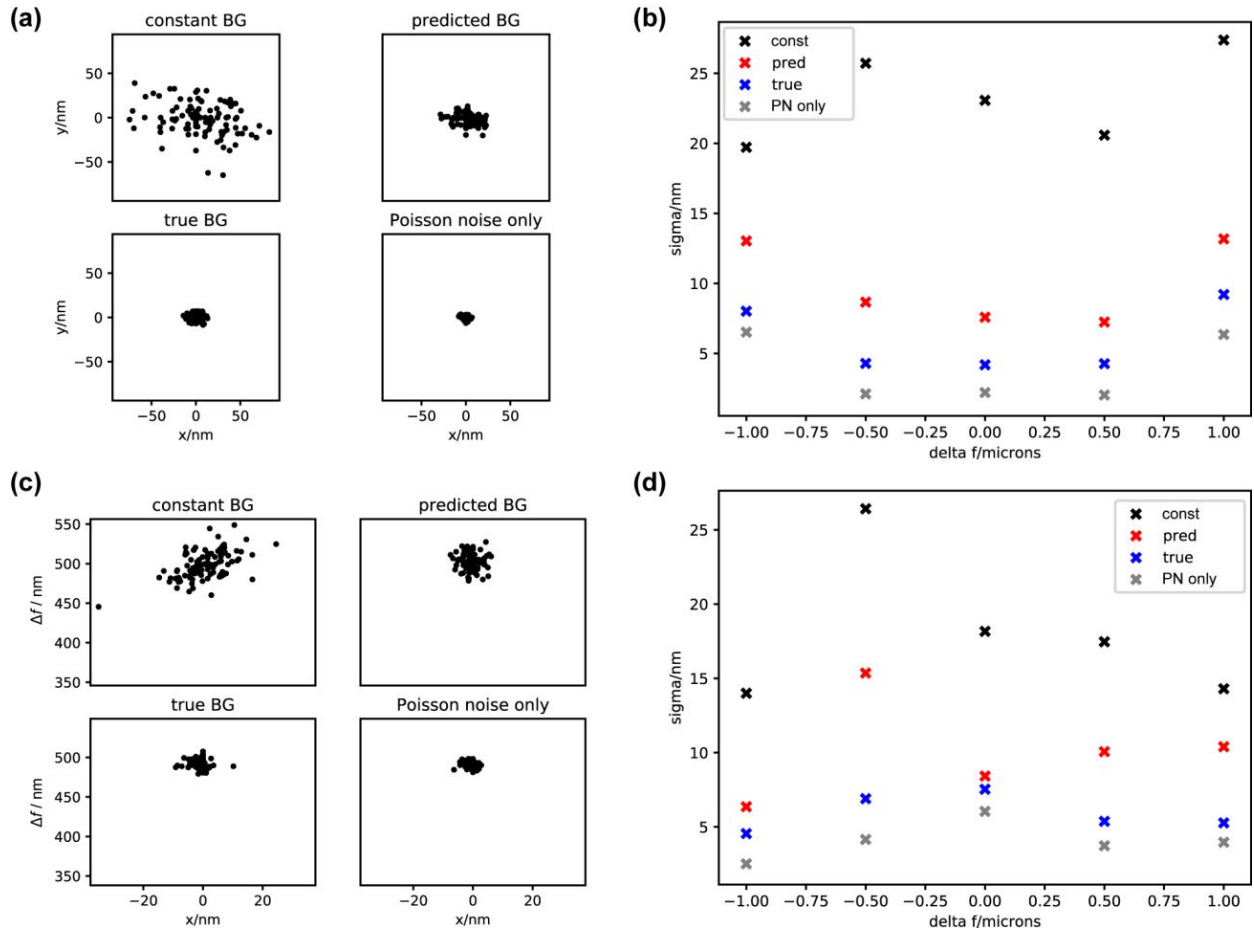

**Figure S8:** Example Scatter plots and corresponding standard deviations for MLE analysis of the arbitrary PSF. (a) and (b) 15,500 signal photons, 150 average BG photons. Emitter at 0 μm, focal position for scatter plot at -0.5 μm. (c), (d) 25,000 signal photons, 87.5 average BG photons. Emitter at 2 μm, focal position for scatter plot at 0.5 μm.

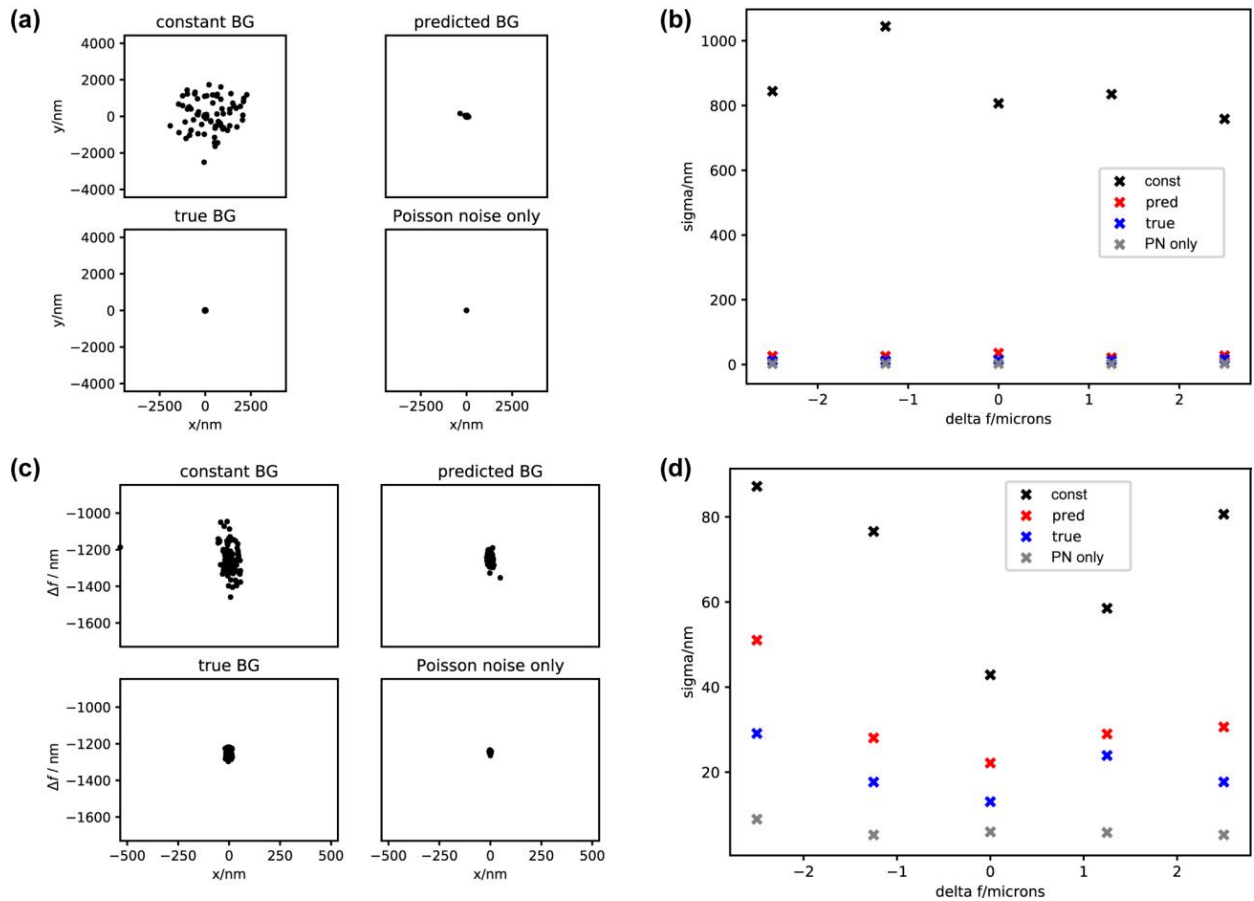

**Figure S9:** Example Scatter plots and corresponding standard deviations for MLE analysis of the Tetra6 PSF. (a) and (b) 10,000 signal photons, 150 average BG photons. Emitter at 0 μm, focal position for scatter plot at 0 μm. (c), (d) 15,000 signal photons, 87.5 average BG photons. Emitter at 10 μm, focal position for scatter plot at -1.25 μm.

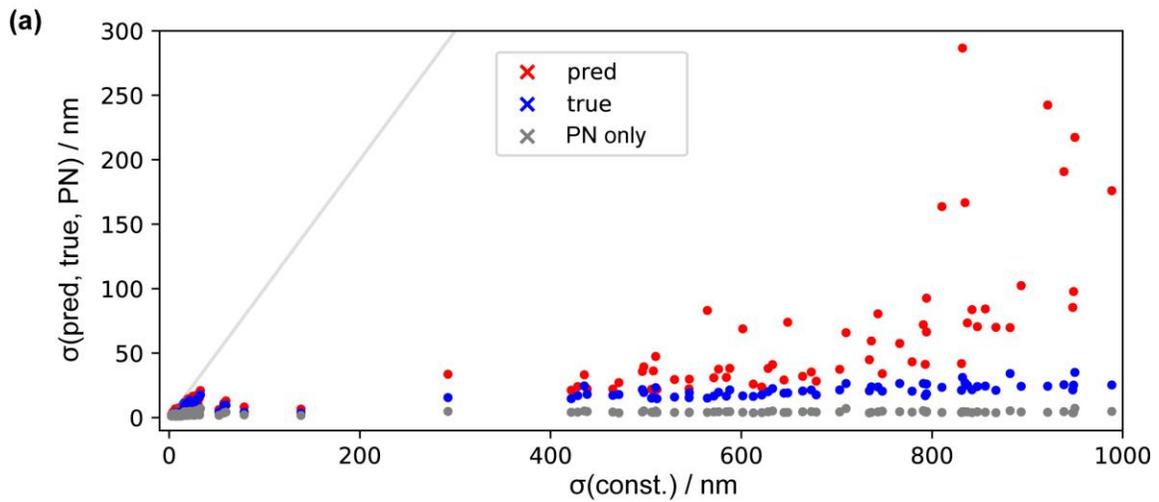

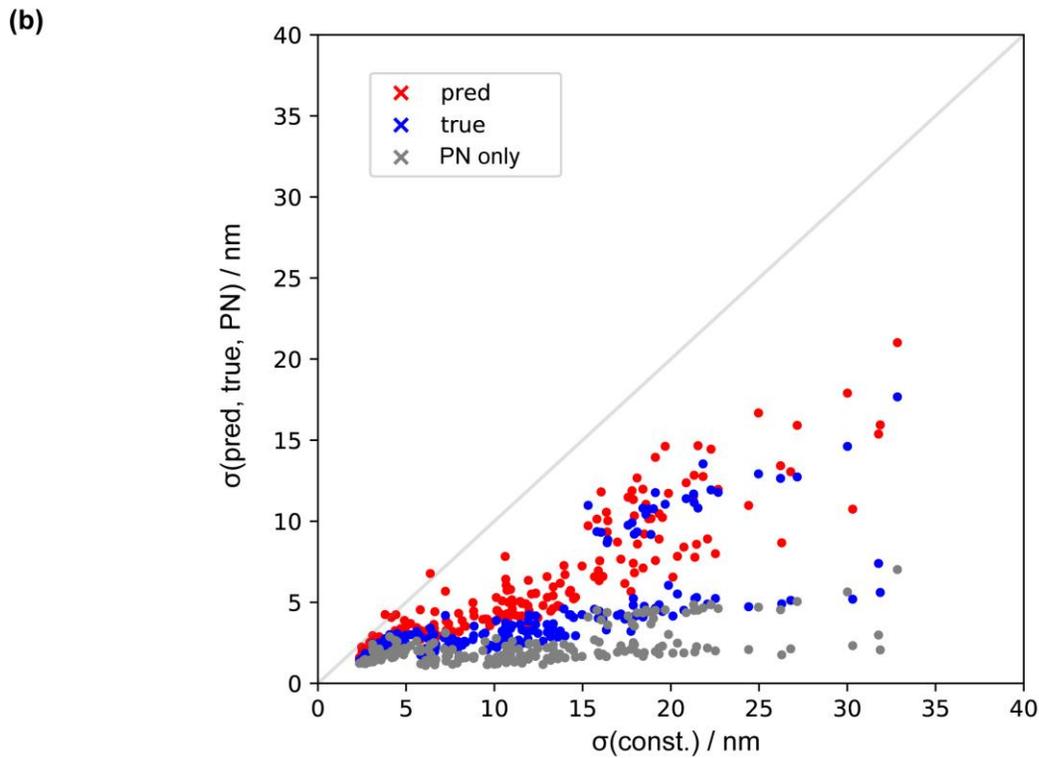

**Figure S10:** Comparison of standard deviations of position estimates for the Tetra6 PSF, 7,500 to 75,000 signal photons. Standard deviations of position estimates for BG correction with the predicted BG or with the true BG and for the Poisson noise-only PSFs are plotted against the standard deviations of position estimates for BG correction with constant BG (analogous to Figure 3). All analyzed conditions are shown. The gray line indicates equal performance. (b) is a zoom-in of (a).

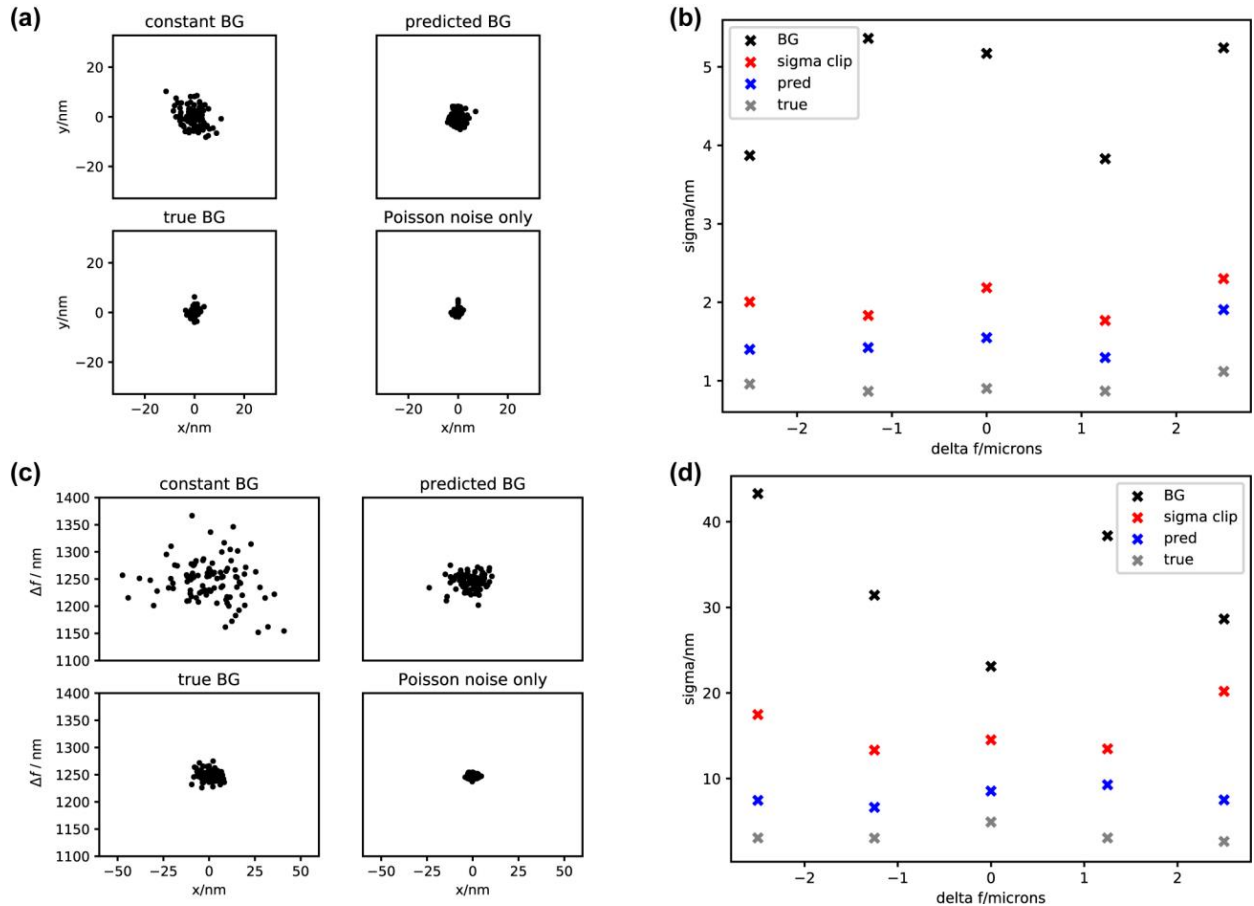

**Figure S11:** Example Scatter plots and corresponding standard deviations for MLE analysis of the Tetra6 PSF, 7,500 to 75,000 signal photons. (a) and (b) 75,000 signal photons, 87.5 average BG photons. Emitter at 2 µm, focal position for scatter plot at -2.5 µm. (c), (d) 41,250 signal photons, 150 average BG photons. Emitter at 6 µm, focal position for scatter plot at 1.25 µm.

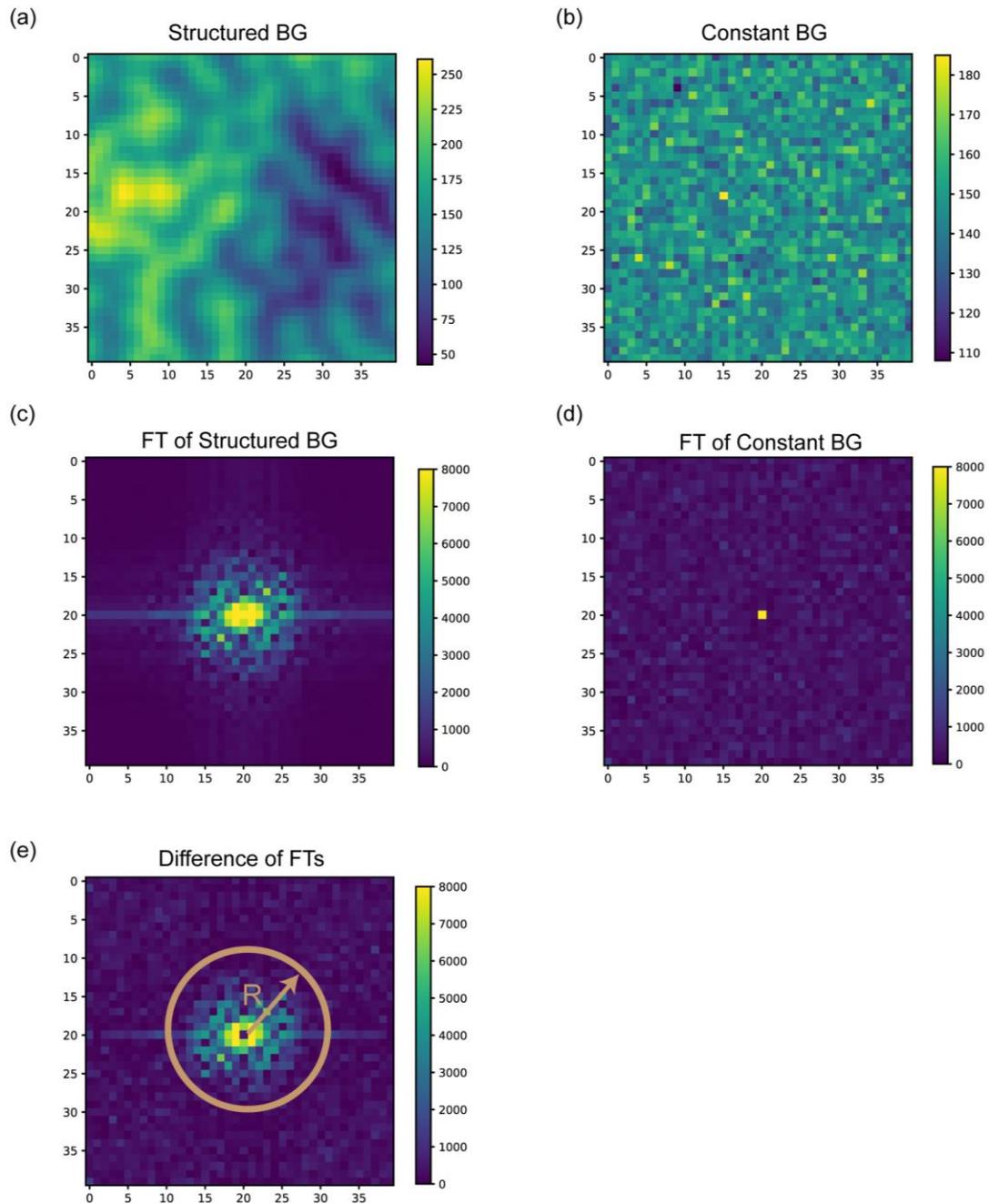

**Figure S12:** Calculation of BG complexity. (a) Representative sBG to be analyzed. (b) Corresponding constant BG with same average photon count per pixel and Poisson noise. (c) Fourier transform of (a). (d) Fourier transform of (d). (e) To remove the dominant lowest spatial frequency, the difference of (c) and (d) is calculated. Next, the integrated weighted radial distribution is determined and normalized by the signal-to-background ratio, yielding the BG complexity.

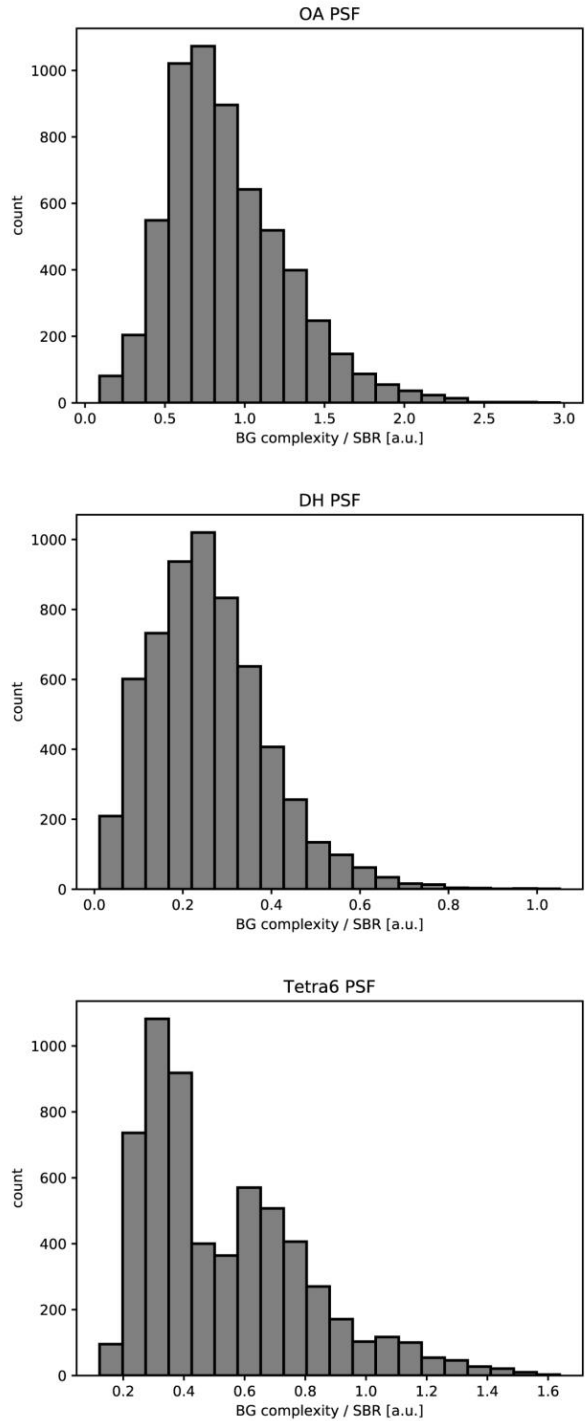

**Figure S13:** Histogram of background complexities for the OA, the DH, and the Tetra6 PSF. Data for all beads and all frames is shown (6x1000 frames for each PSF). Note that the background complexities are sometimes above 1, exceeding the training range of BGnet, which still yielded good results. This highlights the robustness of the method. The BG complexity metric is scaled is in main text Figure 4.

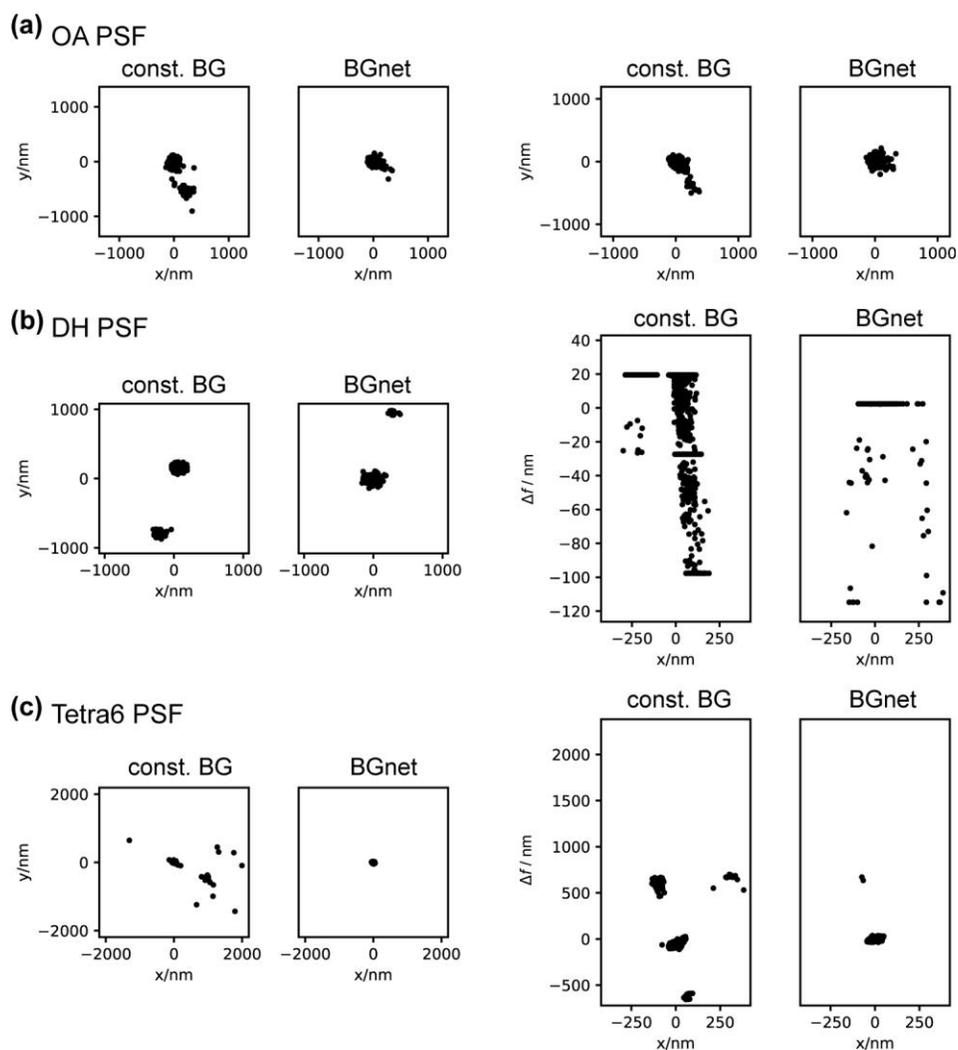

**Figure S14:** Example scatter plots for analyses of experimental PSFs. (a) OA PSF, (b) DH PSF, (c) Tetra6 PSF. Note that the clusters of poorly localized molecules (e.g. x/y scatter for the DH PSF, x/Δf scatter for the Tetra6 PSF) contain only very few localizations in case of BGnet (17 and 2, respectively), whereas they contain hundreds for BG correction with constant BG. The biasing of localizations towards few clusters is typical for MLE fitting when sBG is present. The spread within each individual cluster can be quite low, however, as the effect of the experimental sBG onto the MLE fitting is unpredictable, the induced bias cannot be corrected.

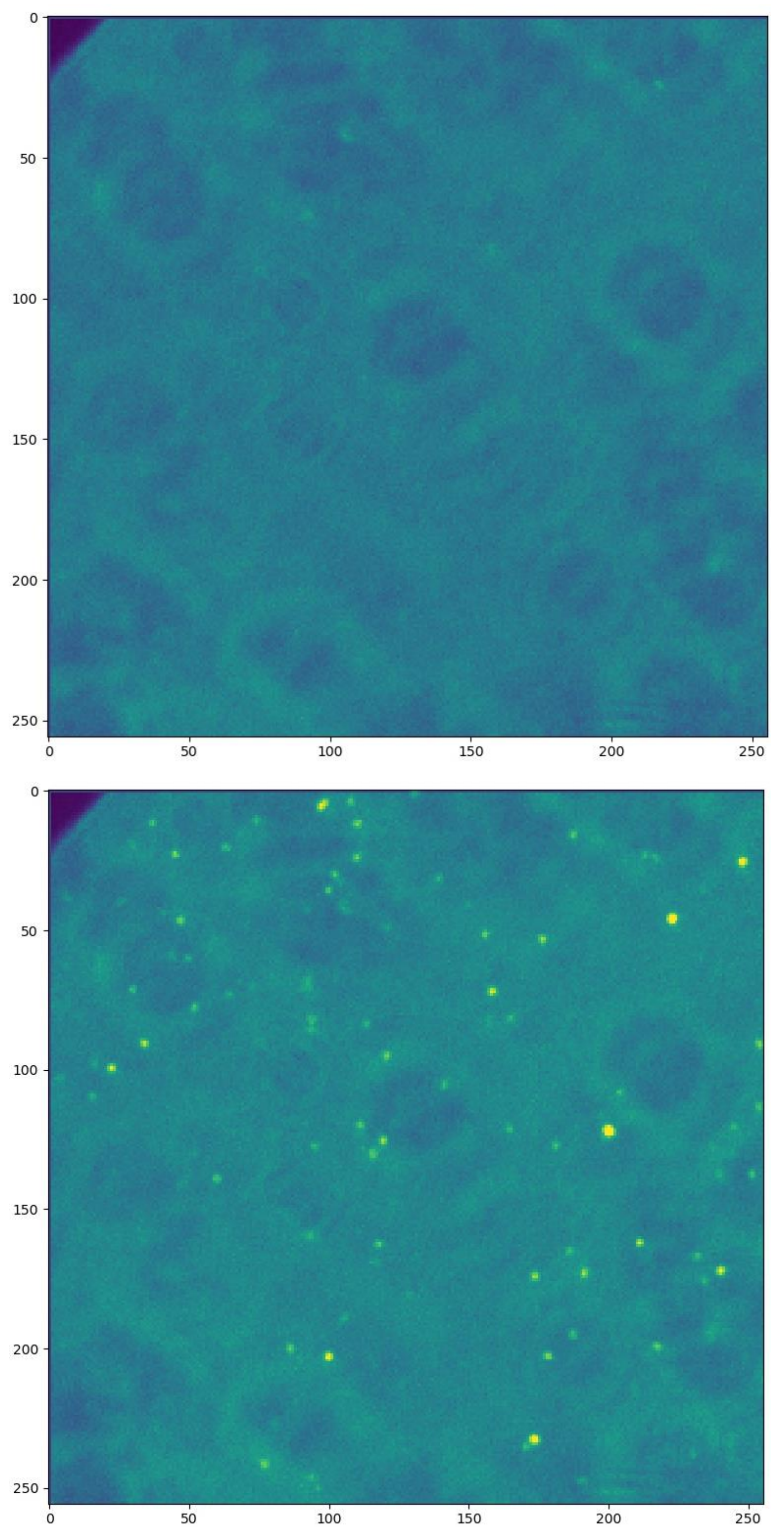

**Figure S15:** sBG that was added to the experimental data (top) and representative frame with single molecules (bottom). Contrast settings are equal between the two images.